\documentclass[fleqn,usenatbib]{mnras}
\usepackage{amsmath,amssymb,graphicx}
\usepackage{newtxtext,newtxmath}
\usepackage[T1]{fontenc}

\DeclareRobustCommand{\VAN}[3]{#2}
\let\VANthebibliography\thebibliography
\def\thebibliography{\DeclareRobustCommand{\VAN}[3]{##3}\VANthebibliography}

\usepackage{microtype}
\usepackage{enumerate}
\usepackage{xcolor}

\title[Evolution of the S0 fraction]{The redshift evolution of the S0 fraction for $z<1$ in COSMOS}

\author[M. K. Cavanagh et al.]{
Mitchell K. Cavanagh,$^{1}$\thanks{E-mail: mitchell.cavanagh@icrar.org (MKC)}
Kenji Bekki and $^{1}$
Brent A. Groves$^{1,2}$
\\
$^{1}$International Centre for Radio Astronomy Research, The University of Western Australia, 7 Fairway, Crawley, WA 6009, Australia\\
$^{2}$Research School of Astronomy and Astrophysics, Australian National University, Mt Stromlo Observatory, Weston Creek, ACT 2611, Australia\\
}

\date{Accepted XXX. Received YYY; in original form ZZZ}

\pubyear{2015}

\begin{document}
\label{firstpage}
\pagerange{\pageref{firstpage}--\pageref{lastpage}}
\maketitle

\begin{abstract}
Lenticular (S0) galaxies are galaxies that exhibit a bulge and disk component, yet lack any clear spiral features. With features considered intermediary between spirals and ellipticals, S0s have been proposed to be a transitional morphology, however their exact origin and nature is still debated. In this work, we study the redshift evolution of the S0 fraction out to $z \sim 1$ using deep learning to classify F814W ($i$-band) HST-ACS images of 85,378 galaxies in the Cosmological Evolution Survey (COSMOS). We classify galaxies into four morphological categories: elliptical (E), S0, spiral (Sp), and irregular/miscellaneous (IrrM). Our deep learning models, initially trained to classify SDSS images with known morphologies, have been successfully adapted to classify high-redshift COSMOS images via transfer learning and data augmentation, enabling us to classify S0s with superior accuracy. We find that there is an increase in the fraction of S0 galaxies with decreasing redshift, along with a corresponding reduction in the fraction of spirals. We find a bimodality in the mass distribution of our classified S0s, from which we find two separate S0s populations: high-mass S0s, which are mostly red and quiescent; and low-mass S0s, which are generally bluer and include both passive and star-forming S0s, the latter of which cannot solely be explained via the faded spiral formation pathway. We also find that the S0 fraction in high-mass galaxies begins rising at higher $z$ than in low-mass galaxies, implying that high-mass S0s evolved earlier.
\end{abstract}

\begin{keywords}
galaxies: evolution -- galaxies: elliptical and lenticular, cD -- galaxies: general
\end{keywords}


\section{Introduction}

Understanding how galaxy morphologies vary over cosmic timescales is a crucial first step towards understanding the physical drivers of galaxy formation and evolution \citep{dressler1980, larson1980, bundy2005, papovich2005, driver2009, ilbert2010, kovac2010, oesch2010, conselice2014, ferreira2022}. Observed galaxy populations, past and present, are also important constraints for cosmological simulations \citep{guo2011, somerville2015, vogelsberger2020}. Of key interest is the nature and evolution of lenticular (S0) galaxies, a galaxy morphology that has long been established as an intermediate type between ellipticals and spirals \citep{hubble1936}. The prevalence of lenticular galaxies, and the manner in which this fraction varies across both cosmic time and in different environments, is important to investigating their main formation pathways, and the physical processes that govern their subsequent evolution.

Multiple mechanisms and physical processes have been proposed to explain the formation of S0s \citep{deeley2020}. These can be grouped into two main pathways. The first of these is the "faded spiral" pathway, which involves the transition from a spiral to a S0 galaxy \citep{dressler1980,fasano2000, poggianti2001,moran2007,laurikainen2010,johnston2014,donofrio2015,bait2017,rizzo2018}.
This is believed to be dominant formation pathway in dense environments \citep{coccato2022}, where there are many supporting factors such as the influence of tidal interactions and mergers \citep{bekki2011,deger2018}, ram pressure stripping \citep{gunn1972} and gas starvation \citep{larson1980,bekki2002}. All of these are capable of stripping gas from spirals and quenching further star formation activity \citep{barr2007}. Similarly, in isolated, low density environments, S0s can naturally emerge via the faded spiral pathway as a result of the passive consumption, and eventual depletion, of in-situ gas needed to sustain star formation \citep{rizzo2018}.

The second major pathway involves the formation of S0s as a result of mergers. Simulations have demonstrated that S0s can emerge as the remnants of major merging \citep{bekki1998,borlaff2014,querejeta2015,tapia2017,eliche-moral2018}, and there is observational evidence in support of major merging as a significant factor in the evolution of massive, early type galaxies \citep{prieto2013, robotham2014}, which agrees well with hierarchical models of galaxy formation and evolution \citep{somerville2015}. In addition, several other mechanisms have been proposed to account for S0 formation, including accretion through successive minor merging \citep{diaz2018}, AGN activity \citep{vandenbergh2009}, disc instabilities \citep{saha2018}, as well as through internal secular evolution and bulge growth \citep{kormendy2004,laurikainen2006,laurikainen2010}. Stellar mass is also believed to play a key role in S0 formation, with studies suggesting that there is a difference in formation pathways between low-mass S0s, which are mostly faded spirals, and high-mass S0s, {which instead likely formed via a different pathway, such as merging \citep{kannappan2009, bellstedt2017, fraser-mckelvie2018, dominguezsanchez2020}.}

It is well established that S0s are more prevalent in dense environments, whereby the fraction of S0s typically increases at the expense of a decreasing fraction of spirals \citep{dressler1980,fasano2000,cappellari2011,mishra2019}. However, the extent to which the S0 fraction varies over redshift is less certain. Some studies have posited that there has been little to no evolution of the relative fraction of S0s with redshift \citep{vanderwel2007,holden2009,vulcani2011}, while others have found evidence for an increasing S0 fraction with decreasing $z$ in both cluster environments \citep{dressler1997, desai2007,poggianti2009,just2010}, as well as in the field \citep{oesch2010, huertas-company2015}. 

Studies into the evolution of galaxy morphology are contingent on the availability of classifications. However, the classification of galaxies by visual inspection is a slow, time-consuming task. Recent citizen science efforts have been vital in enabling visual classification to effectively scale to larger datasets \citep{willett2013,masters2021,walmsley2021}, however the speed of the classification remains unchanged. As such, automated classification of galaxy morphologies is paramount in enabling vast volumes of data to be analysed quickly and efficiently, especially data at scales that render visual classification infeasible \citep{beck2018}. Deep learning models, such convolutional neural networks (CNNs) \citep{lecun2015}, have been utilised to rapidly and efficiently analyse images of nearby, low-redshift galaxies with great success. Recent applications not only include morphological classification \citep{dieleman2015, dominguezsanchez2018, cavanagh2021, vega-ferrero2021}, but also include similarity-based clustering \citep{martin2020, cheng2021, walmsley2022}, as well as morphological segmentation \citep{hausen2020}. The advantage of a deep learning approach lies in its efficiency and versatility, the latter of which is especially pertinent for CNNs, which are especially adept at classifying images via feature extraction \citep{simonyan2014}. While the majority of studies employing CNNs for classification have focused on readily available datasets of nearby, low-redshift galaxies, such as Galaxy Zoo \citep{willett2013}, 
fewer studies have used a deep learning approach to classify high-redshift galaxies. The morphological classification of high-redshift galaxies is far more challenging due to the inherent limitations in image quality, and the lack of any sufficient large-scale dataset of known morphologies with which to train a CNN.

The Cosmic Evolution Survey (COSMOS) \citep{scoville2007} is a deep, multi-wavelength survey aimed at investigating the evolution of galaxies. In this work, we develop and adapt several deep learning classifiers, initially trained on $g$-band SDSS images of galaxies from the \citet{nair2010} morphological catalogue (hereafter NA10), to classify COSMOS images via transfer learning. {Transfer learning refers to a family of machine learning techniques designed to leverage the pre-existing capabilities of an initial, pretrained model in order to perform some task in a related domain (see \citealt{weiss2016} for a review). Recent studies in astronomy have used this technique to adapt models trained on simulated galaxies to instead classify real galaxies \citep{ghosh2020} (and vice-versa; \citealt{cavanagh2022}), in addition to adapting models to classify images from different surveys \citep{dominguezsanchez2018}.}

The images we classify are $i$-band images from the Hubble Space Telescope Advanced Camera for Surveys (HST/ACS) \citep{koekemoer2007,massey2010}. Although COSMOS extends to redshifts beyond $z > 3$, we limit our analysis to samples with $z < 1$. This is primarily to ensure that the images are at a reasonable resolution suitable for classification with our CNNs, and also to limit the adverse effects of $k$-correction and PSF-smoothing. Despite the HST imaging being deeper in surface brightness, the high-redshift COSMOS galaxies appear significantly noisier than the low-redshift SDSS galaxies, which is a barrier to any direct application of a deep learning model. We overcome this barrier through the use of data augmentation to augment our existing SDSS images with artificial noise such that they mirror the characteristic noise and quality of the COSMOS images. We then use these augmented images to fine-tune our initial models. The key advantage of this approach is that the classifications utilise the existing known morphologies, which is significant since these include a specific category for S0s. We find that this transfer learning approach allows us to classify S0s with a significantly higher accuracy compared to an otherwise direct application.

The structure of our paper is as follows. In \S 2, we briefly summarise our sample selection process, and how these images were then preprocessed and classified by our models. We further describe our CNN model, including its architecture, and also describe the data augmentation and transfer learning procedures. Lastly, we describe how we obtain final classifications for each of the COSMOS images. In \S 3, we present our results on morphological classifications and the redshift evolution of each morphology, focusing on the growth in the S0 fraction. We discuss the physical properties of our classified S0s, and show that they comprise two distinct populations. We also provide some example classifications, including those on which our 3-class and 4-class models disagree. In \S 4, we discuss our results and implications for the formation and evolution of S0s. We further evaluate our deep learning approach and use of data augmentation. We also examine the model's performance on our NA10 test set not just for comparison, but also to judge the impact of the noise augmentation. Finally, we conclude this paper with a summary of our key results in \S 5.

\section{Methods}

\subsection{Datasets}

\begin{figure}
\centering
\includegraphics[scale=0.59]{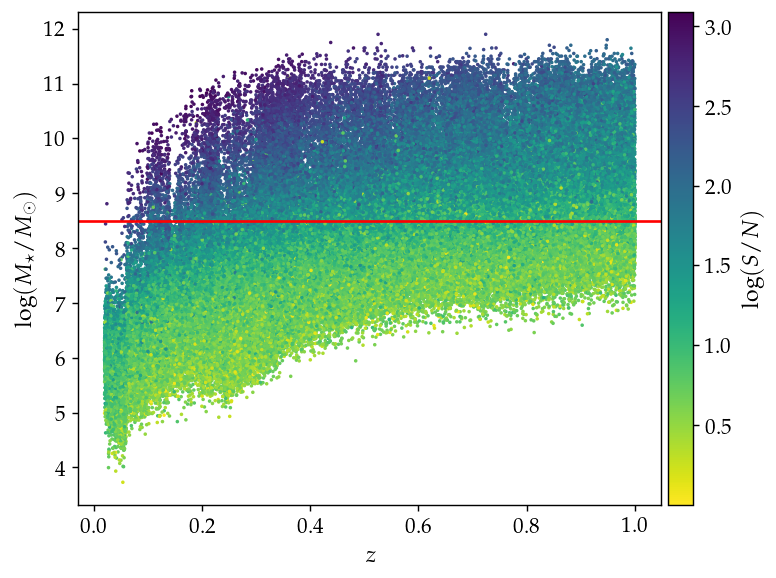}
\caption{Full COSMOS sample for redshifts $z < 1$ with redshift on the $x$-axis and stellar mass on the $y$-axis. Samples are shaded according to their signal-to-noise ratio $\log(\text{S}/\text{N})$. The solid, red line denotes the minimum mass considered for this study.}
\label{fig:sample-selection}
\end{figure}

\subsubsection{COSMOS Sample Selection}

The sample of galaxies used for this work consists of a subsample of galaxies from the COSMOS2020 catalogue \citep{weaver2022}. We select all galaxies with $z < 1$ using the \textsc{LePhare} best redshift \citep{arnouts2011}, and apply a lower mass cut of $\log(M_\star/M_\odot) = 8.5$ with no upper limit. Figure \ref{fig:sample-selection} displays the mass distribution (\textsc{LePhare} best mass) of the $z < 1$ COSMOS2020 sample, illustrating our lower mass limit. Samples are coloured according to signal-to-noise, which is defined as the ratio of F814W flux to flux error \citep{leauthaud2007}. Figure \ref{fig:sample-selection} illustrates the importance of a lower mass limit, which is necessary to discard samples with poor signal-to-noise. As a result, our sample subsequently consists of galaxies with $\log(S/N) > 1$. Importantly, we note that there are relatively few high-mass galaxies at redshifts below $z \approx 0.2$. Furthermore, signal-to-noise degrades with redshift, which has the greatest impact on low-mass galaxies.

\begin{figure}
\centering
\includegraphics[scale=0.52]{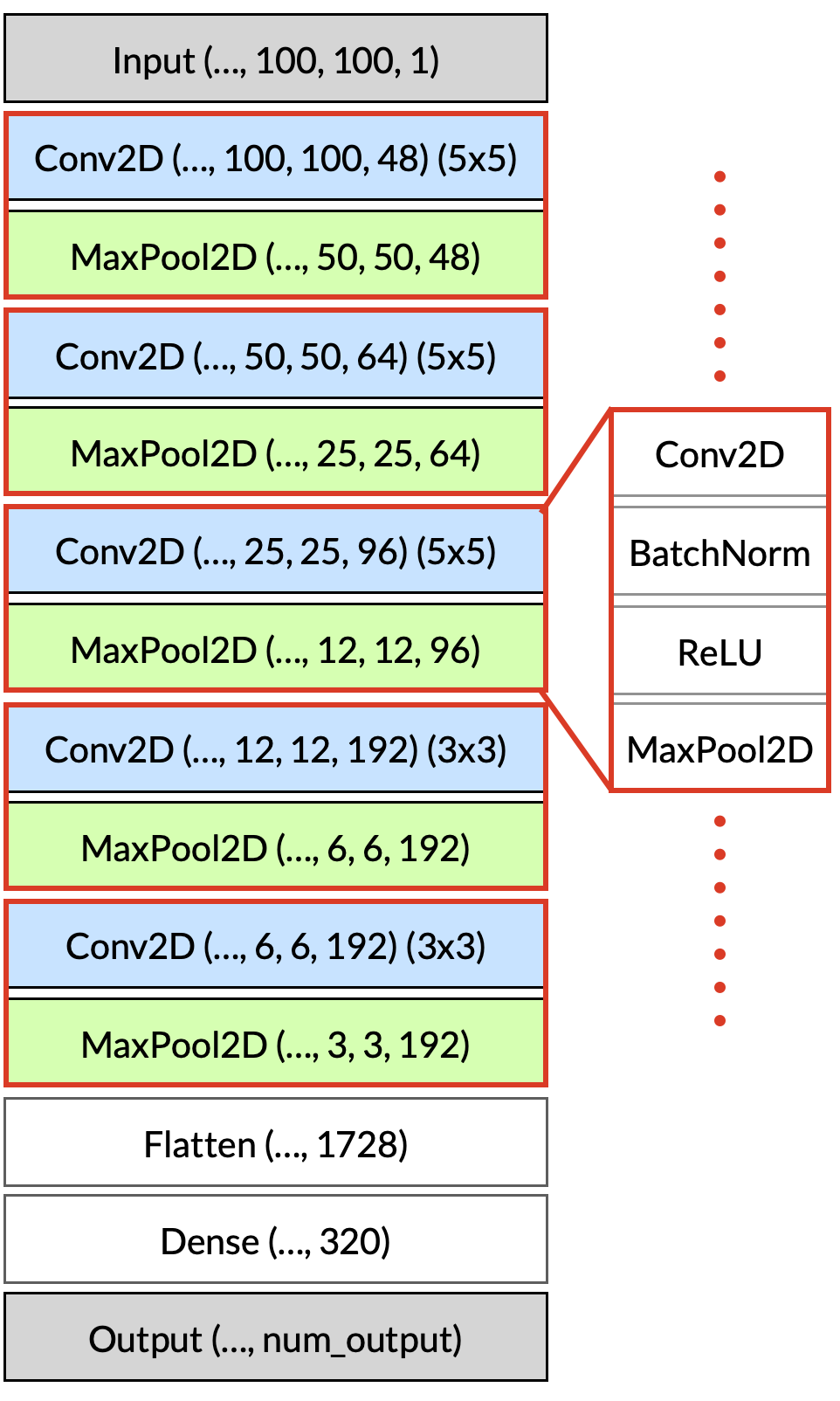}
\caption{Outline of our convolutional neural network model. The tuples in parentheses indicate the output shape of each layer. Ellipses ($\ldots$) denote the batch size. There are five blocks (shown with red borders) of alternating 2D convolution (\texttt{Conv2D}) and 2D max pooling (\texttt{MaxPool2D}) layers. These have output shapes of the form (\texttt{batch\_size}, \texttt{width}, \texttt{height}, \texttt{channels}), where \texttt{channels} denotes the number of feature maps (and hence the number of convolutional filters). Furthermore, the kernel size for each \texttt{Conv2D} layer is shown in parentheses to the right of the output shape. Each convolution is immediately followed by both a batch normalisation layer, and then a ReLU activation layer (see the expanded block). In total, the model contains 1.28 million trainable parameters.}
\label{fig:cnn-schematic}
\end{figure}

Having established our dataset, we then obtained HST-ACS \citep{koekemoer2007} image cutouts for each sample from the NASA/IPAC Infrared Science Archive \footnote{\url{https://irsa.ipac.caltech.edu/data/COSMOS/overview.html}}. These are monochrome $i$-band (F814W filter) images with a pixel scale of 0.03''. All cutouts were obtained with an angular size corresponding to the same physical scale of 50 kpc. This is to match the scale of the SDSS images used to train the original model (see next subsection for details). Consequently, the angular sizes of the cutouts vary with redshift, resulting in images with a range of pixel sizes. We calculated the desired angular diameter in pixels for each sample with redshift $z$ using \textsc{Astropy} \citep{theastropycollaboration2013}, in particular the \texttt{cosmology} package using the cosmological parameters from WMAP9 \citep{hinshaw2013}. Not all samples in our subsample returned a cutout from the image server, however we were able to obtain a total of 85,378 images; corresponding to about 94\% of the original subsample. All the images were then resized to a size of 100x100 pixels using the \textsc{Pillow} package \citep{vankemenade2022} { using the default bi-cubic interpolation setting}, and then linearly normalised so that each pixel in each cutout has a value strictly between 0 and 1 inclusive, where 1 is the brightest pixel. This is the final size and format of the images prior to being inputted to the convolutional neural network.

\subsubsection{The Training Data}

The training data used to train the initial models consists of 14,034 $g$-band DR7 SDSS images \citep{abazajian2009} of galaxies from the \citet{nair2010} (NA10) morphological catalogue. In particular, the catalogue consists of all spectroscopic targets in the DR4 SDSS release \citep{adelman-mccarthy2006} with magnitudes brighter than an extinction-corrected $g$-band mag $=16$ between $z \approx 0.01$ and $z=0.1$. Each image shares the same physical scale of 50kpc$\times$50kpc. Each galaxy is visually classified according to numerical, Hubble T-Types. We group these T-Types together into broader elliptical, lenticular, spiral and irregular/miscellaneous morphological categories (see \citealt{cavanagh2021} for exact details). We do not place any restrictions on the physical sizes of the galaxies themselves. The training data underwent a similar data processing regime as described for the COSMOS sample. We further partitioned the full training data into separate training and test sets according to an 80:20 split. The training set is used to train the model, while the test set is reserved for final evaluation.

\subsection{The CNN Models}

\begin{figure}
\centering
\includegraphics[scale=0.68]{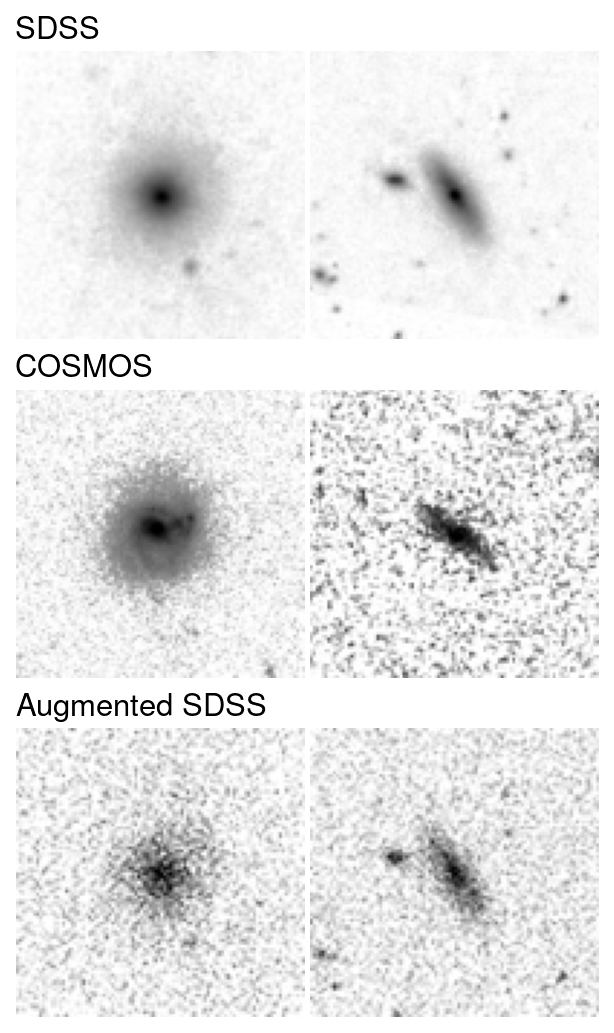}
\caption{Example of our noise augmentation procedure. The top row shows some example SDSS images. The middle row shows some visually similar COSMOS images. The bottom row shows the same SDSS images as in the top row, but this time after having been modified with additional, artificial noise.}
\label{fig:noiseaug-example}
\end{figure}

Convolutional neural networks are specialised neural networks that are especially well-suited to analysing image data due to their capacity for abstract feature extraction. To process the input data and extract relevant features, CNNs use successive convolutional layers. These layers linearly convolve their given inputs with several filters, or kernels. The weights for these filters are initially random, but are progressively tweaked and optimised as the network is trained. It is in this manner that the CNN learns to extract features automatically. The use of learnable filters as opposed to pre-determined, hard-coded filters is one of the key advantages of CNNs, making them ideal models for general-purpose image classification across many disciplines. For a full treatment of CNNs, their operation and their training, see \citet{haykin2009, lecun2015,goodfellow2016}.

The models developed for this study are adapted from our previous work \citep{cavanagh2021} (hereafter C21). {In particular, we develop two models: one designed to classify galaxies as either elliptical, lenticular or spiral (3-class model); and another model which adds a fourth category for irregular/miscellaneous galaxies (4-class model). We train these models on 100x100 pixel sized g-band SDSS images that are each labelled with their known morphologies from the NA10 morphological catalogue. Our models are developed and trained in \textsc{Python} using the \textsc{TensorFlow} \citep{abadi2016} and \textsc{Keras} \citep{chollet2015} libraries. We utilise a similar data processing and training procedure as in C21, however the major difference is that this work utilises a new model architecture, as outlined in Figure \ref{fig:cnn-schematic}. This new architecture underwent extensive hyperparameter tuning using \textsc{Optuna} \citep{akiba2019}, a general-purpose optimisation framework. We find that this new architecture enables higher classification accuracies in both the 3-class and 4-class cases compared to that originally used in C21.} Figure \ref{fig:cnn-schematic} shows some of these optimised hyperparameters including the number of convolutional filters, convolutional kernel sizes and the number of nodes in the dense layer. Our model uses batch normalisation layers \citep{ioffe2015} after each convolutional layer; these layers serve to better regularise the model and improve model robustness. We use ReLU activation functions throughout the model, with softmax activation in the output layer in order to output probabilities for each class. These probabilities are also referred to as classification confidences (or simply confidence) and we will use these terms interchangeably. We use the Adam optimiser \citep{kingma2014} with an initial learning rate of $8 \times 10^{-4}$, and we use categorical cross-entropy as our loss function. The model contains a total of 1.28 million trainable parameters.

\begin{figure}
    \centering
    \includegraphics[scale=0.56]{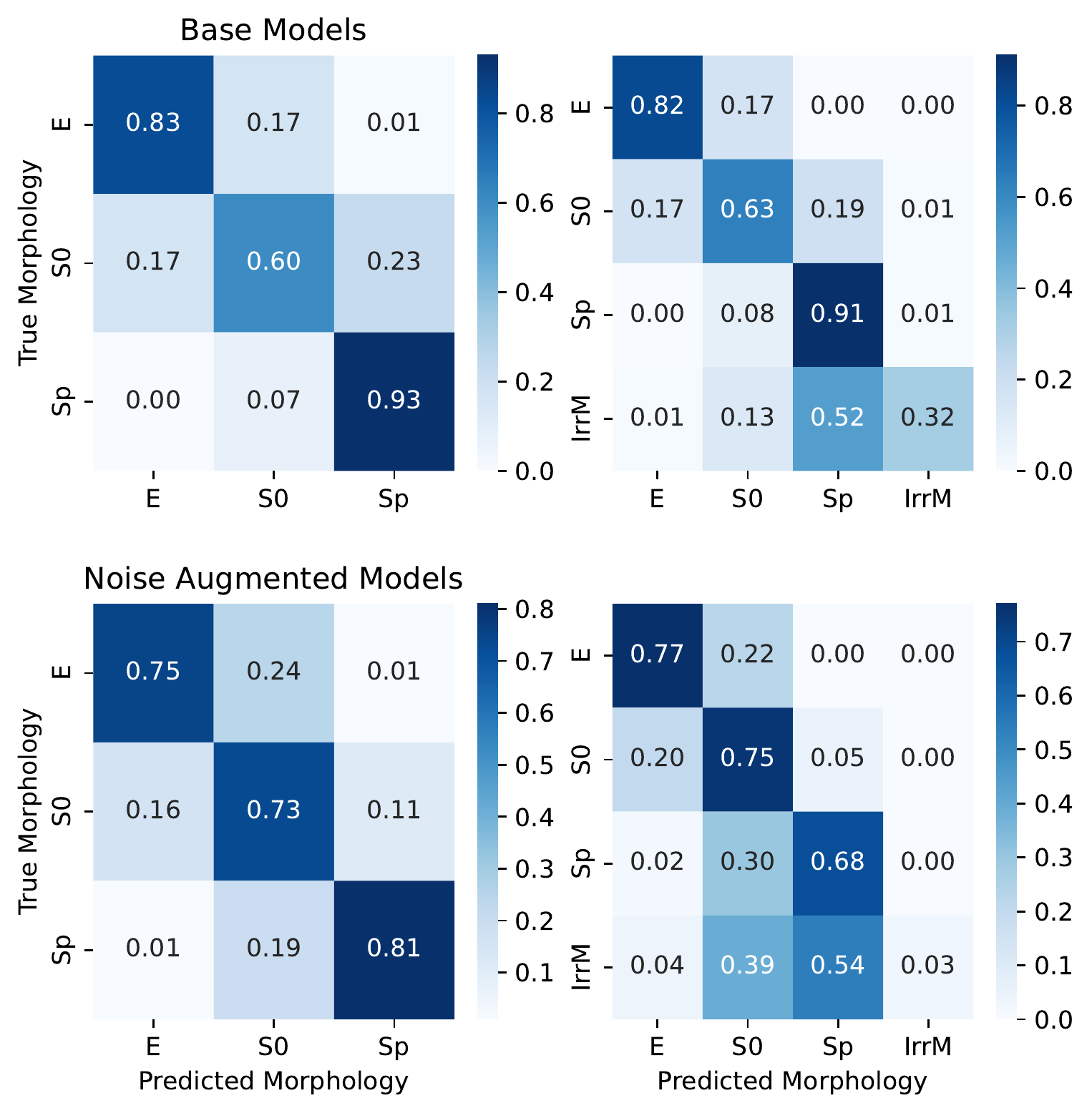}
    \caption{Confusion matrices for the base 3-class and 4-class models (top row), as well as the noise-augmented 3-class and 4-class models (bottom row), based on final evaluation with the holdout NA10 test set.{ Rows denote the true morphologies, while columns denote the predicted morphologies. The entries $i,j$ denote the fractions of samples with true morphology $i$ that are classified as the predicted morphology $j$.} The diagonal entries correspond to recall (a.k.a. per-class accuracy). The same test set is used for both models, with the augmented models tested on augmented images. For the noise augmented models, the predictions (and hence accuracies) are based on the mean confidences of each of the five constituent ensemble models.}
    \label{fig:confmat}
\end{figure}

\subsection{Model Development and Evaluation}

The specific models that we use to classify our sample of COSMOS images are adapted from models that have been pretrained on SDSS images from the NA10 catalogue (these are referred to hereafter as the base models). To perform this model adaptation, we use a combination of transfer learning and extensive data augmentation. In particular, we train new ensembles of 3-class and 4-class models. In each case, the models are initialised with the weights of the pretrained base model. These models are then retrained on SDSS images that have been injected with artificial Gaussian noise. This artificial noise is designed to resemble the characteristic range of noise levels in the images from our COSMOS subsample. The purpose of this data augmentation is to train the models to better classify noisy images. Full technical details regarding the methodology and results of this technique are given in Appendix A, where testing demonstrates that models trained with this technique perform considerably better at classifying noisy images. Due to the high variance associated with the noise augmentation and transfer learning, our final 3-class and 4-class models are model ensembles, each consisting of five independent models. The final classification is thus based on the output classification probabilities averaged across each individual model in the ensemble.

Figure \ref{fig:confmat} shows the confusion matrices for our original base 3-class and 4-class models, as well as for our final, augmented model ensembles. The most immediate impact of the noise augmentation is that the per-class accuracies for the first three categories (E, S0 and Sp) are smoothed. While the elliptical and spiral accuracy falls, the S0 classification accuracy increases from 60\% to 73\% in the 3-class model, and from 63\% to 75\% in the 4-class model. The noise augmentation also results in a notable fall in spiral accuracy, particular in the 4-class model, with up to 30\% of true spirals instead classified as S0s. Curiously, in the converse, only 5\% of true S0s were misclassified as spirals. It is important to keep in mind that the base models had a broad spread in classifications; 17\% of true S0s were misclassified as elliptical, with up to 23\% misclassified as spirals. This spread is reduced for the noise augmented models. Across the board, however, more samples are predicted to be S0s. We discuss these classification accuracies, their implications, and possible reasons for the shift in overall accuracies in further detail in Section 4.

\begin{figure*}
    \centering
    \includegraphics[scale=0.63]{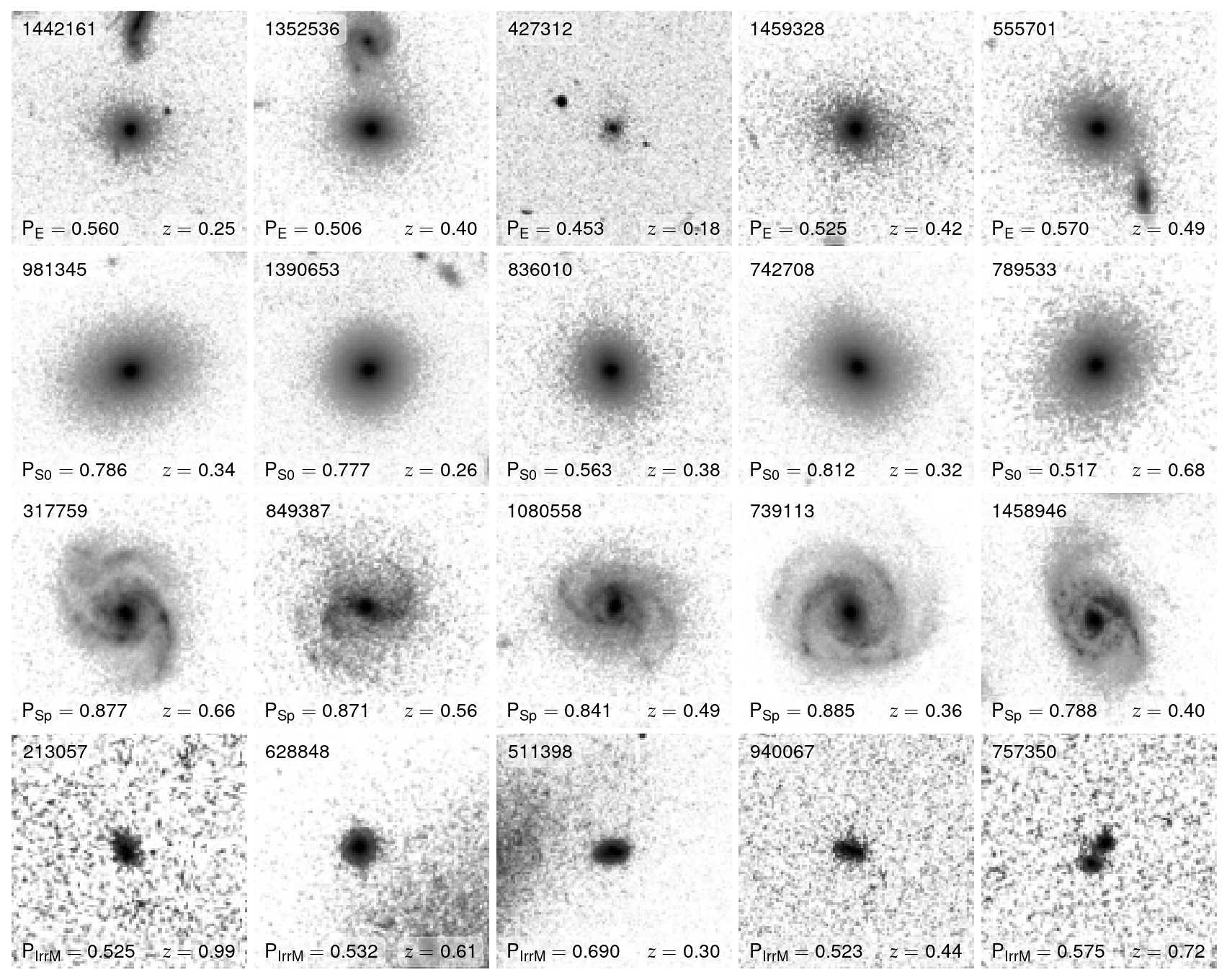}
    \caption{Random selection of COSMOS galaxies classified into each of the four morphological categories by our 4-class model. Each image is annotated with its COSMOS2020 ID (top-left), its classification confidence (bottom-left), and its redshift $z$ (bottom-right).}
    \label{fig:examples_4class}
\end{figure*}

\section{Results}

\subsection{Morphological Classification}

\begin{figure*}
\centering
\includegraphics[scale=0.6]{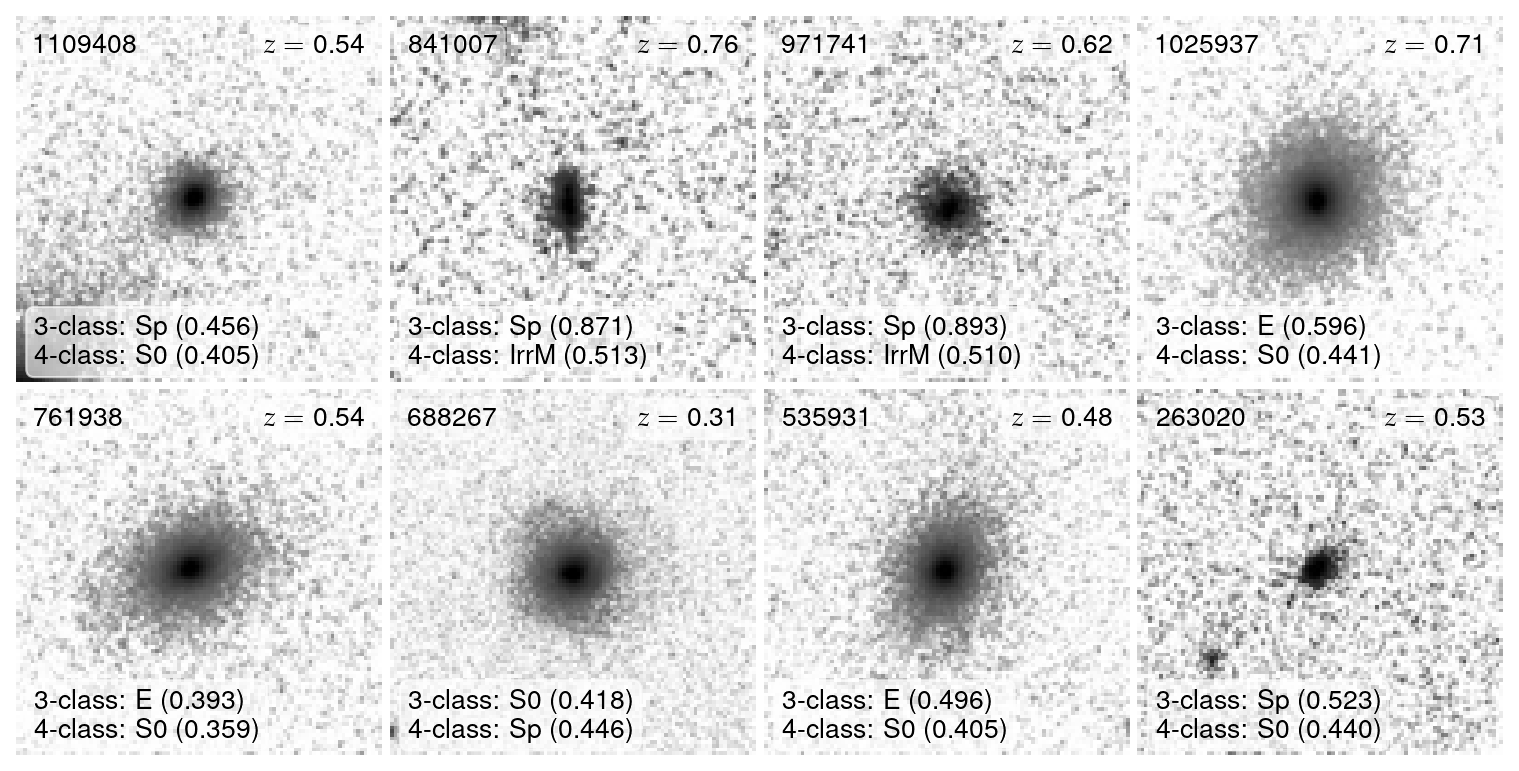}
\caption{Random selection of COSMOS galaxies that have been classified differently by the 3-class and 4-class models.}
\label{fig:examples_differ}
\end{figure*}

We classified all 85,378 images with our noise augmented 3-class and 4-class model ensembles. Since the 4-class model performed the best at classifying S0 galaxies (higher recall), it will be the primary focus of analysis for the remainder of the paper. Figure \ref{fig:examples_4class} shows a random selection of COSMOS galaxies classified into each of the four morphological classes by our 4-class model. Ellipticals are mostly fuzzy, featureless spheroids, while S0s have more pronounced discs and, in general, better defined edges. Spirals all exhibit spiral arms, with ID 849387 in particular illustrating how noise corrupts the appearance of spiral arms. The irregulars/miscellaneous is a broad category that, in the case of the NA10 training data, includes proper irregulars, along with disrupted/interacting galaxies and all other galaxies with unclear morphology. The classified irregulars in COSMOS mainly manifest as small blobs from which it is hard to resolve any clear internal structure. It is likely that some of these, such as ID 628848, may in fact be low mass early type galaxies.

Except for irregulars, the vast majority of galaxies (over 96\%) were assigned the same classification by our 3-class and 4-class model. However, some galaxies were classified differently, with 3-class spiral to 4-class lenticular being the most common classification change. Figure \ref{fig:examples_differ} shows a random selection of galaxies that were given different classifications. Importantly, of those that changed from spiral in the 3-class model to lenticular in the 4-class model (or vice versa), or from elliptical to lenticular, the classification confidences in both models remain low. These are examples of galaxies which the models have difficulty classifying. We will discuss the limitations of our models in more detail in Section 4.3.

\begin{figure}
\centering
\includegraphics[scale=0.61]{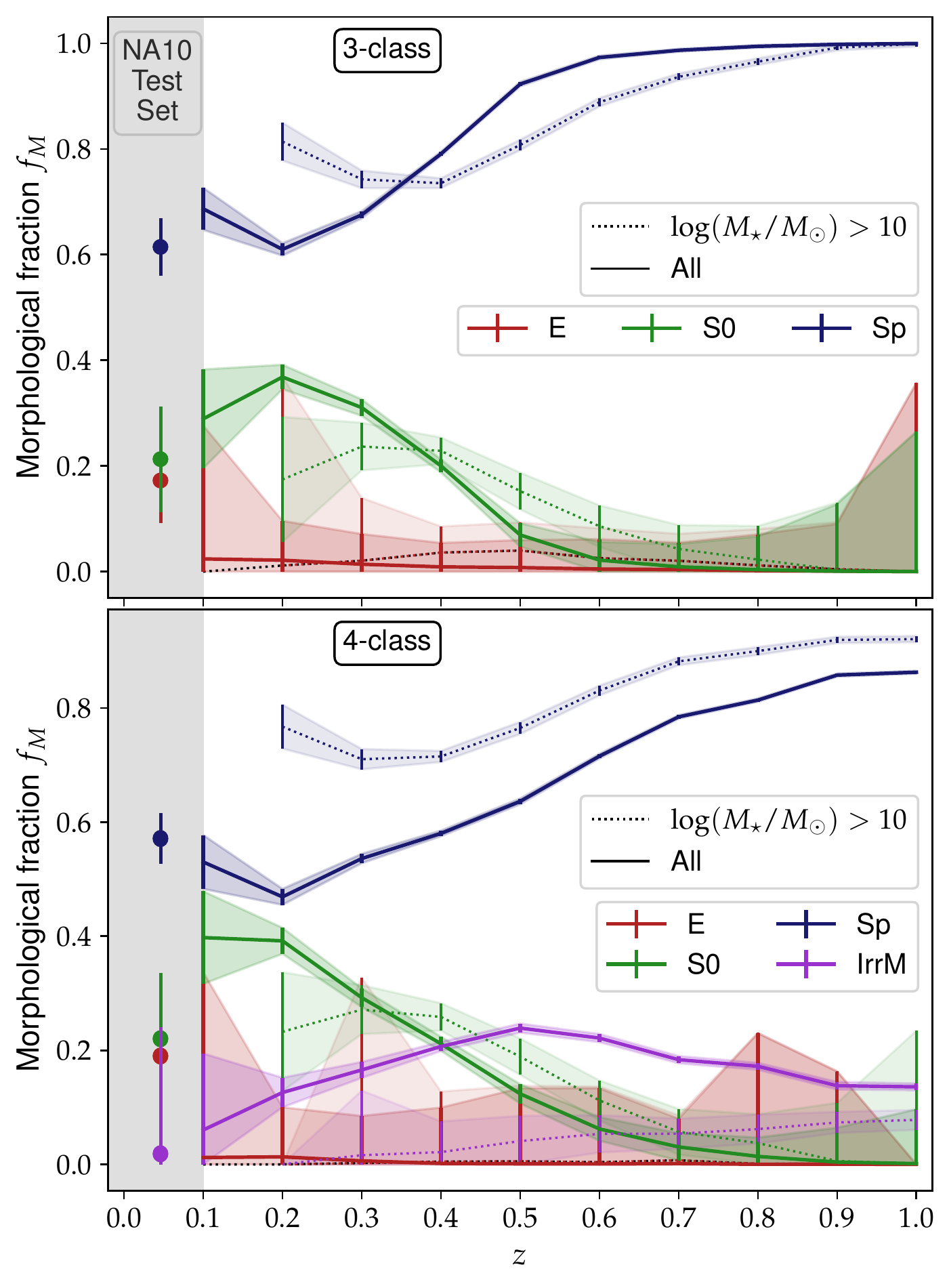}
\caption{{The redshift evolution of the morphological fractions $f_M$ for each morphological class: ellipticals (red), lenticular (green), spiral (blue) and irregular/miscellaneous (mauve). Solid lines indicate the fractions for all galaxies, while the dotted lines denote only the galaxies with stellar masses $\log(M_\odot / M_\star) > 10$. The top panel shows the results from the 3-class ensemble, while the 4-class ensemble results are shown in the bottom panel. Shaded regions denote 1$\sigma$ standard error based on the classification accuracies for each class.}}
\label{fig:ensemble_fracs}
\end{figure}
\begin{figure}
\centering
\includegraphics[scale=0.58]{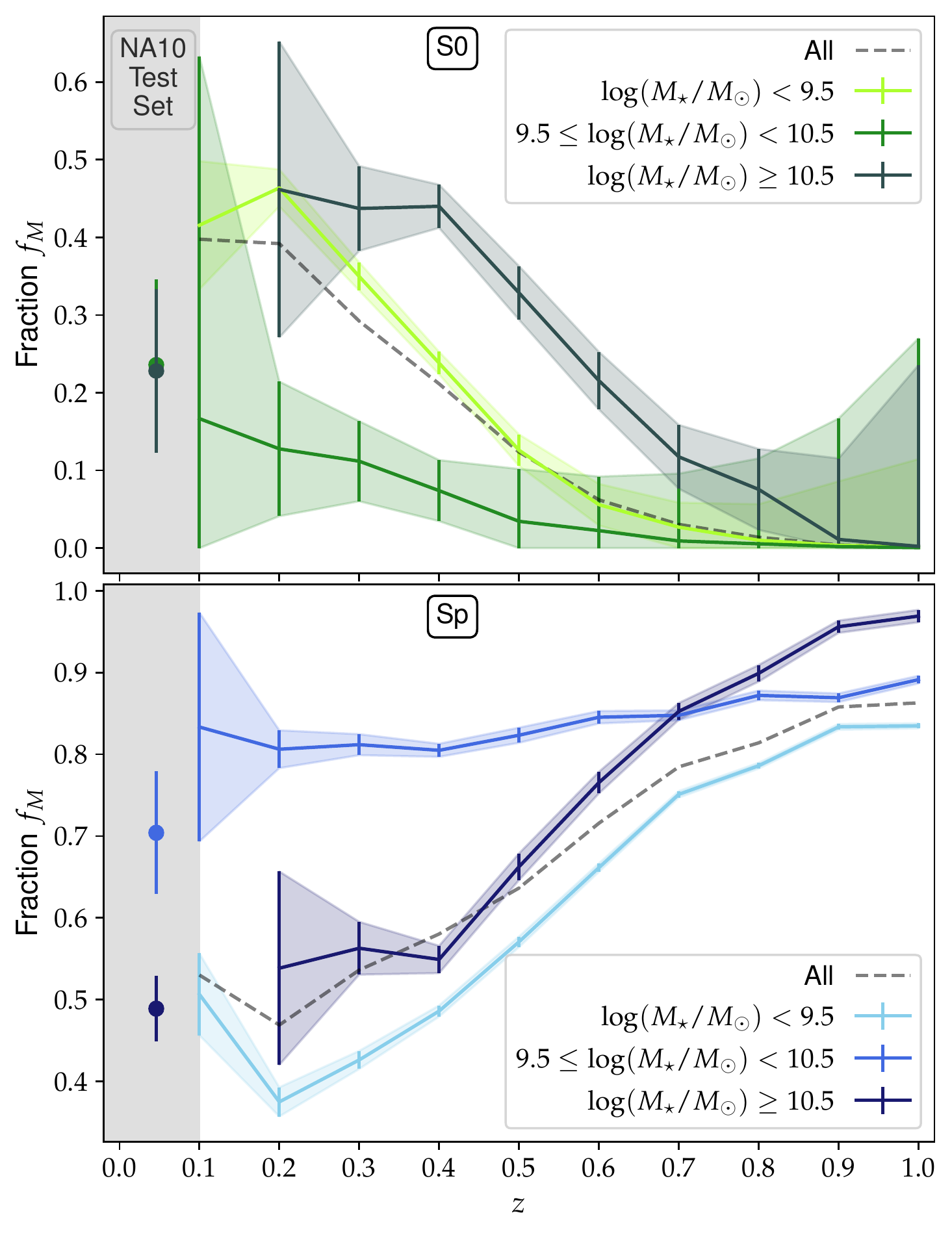}
\caption{{The redshift evolution of the morphological fractions $f_M$ of S0 (top) and Sp (bottom) galaxies in the low mass, intermediate mass and high mass ranges. Fractions are based on the classifications of the 4-class ensemble models.}}
\label{fig:s0fraction_bymass}
\end{figure}

\subsection{Morphological Fractions}

\begin{figure*}
\centering
\includegraphics[scale=0.54]{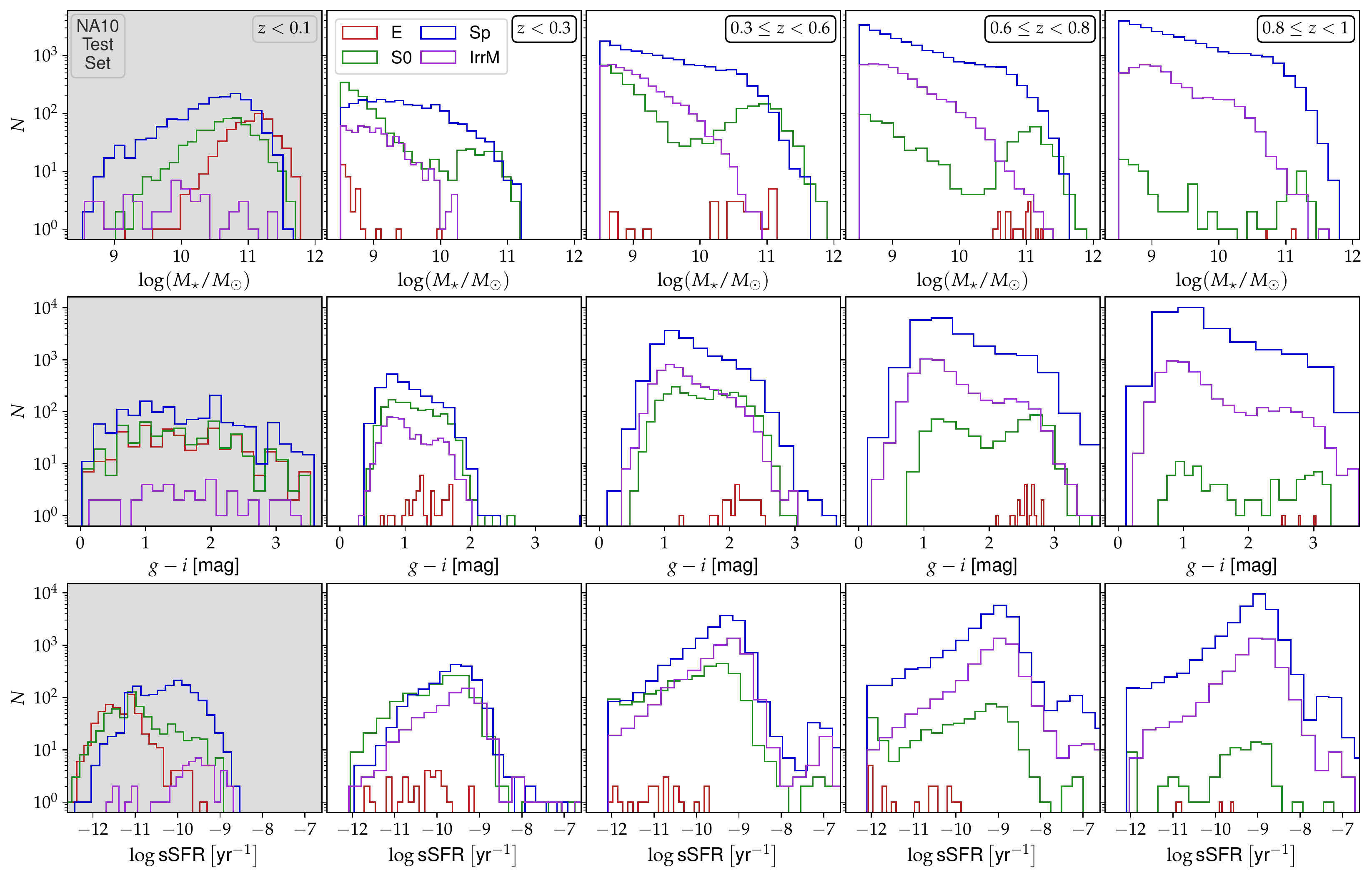}
\caption{{Histograms showing the distributions of stellar mass $\log(M_\star/M_\odot)$ (top row), colour $g-i$ (middle row) and specific star formation rate (sSFR) (bottom row), for samples classified as elliptical, S0, spiral and irregular over different redshift ranges. The leftmost, shaded column shows the corresponding distributions for samples in the NA10 test set.}}
\label{fig:prophist}
\end{figure*}

\begin{figure*}
\centering
\includegraphics[scale=0.61]{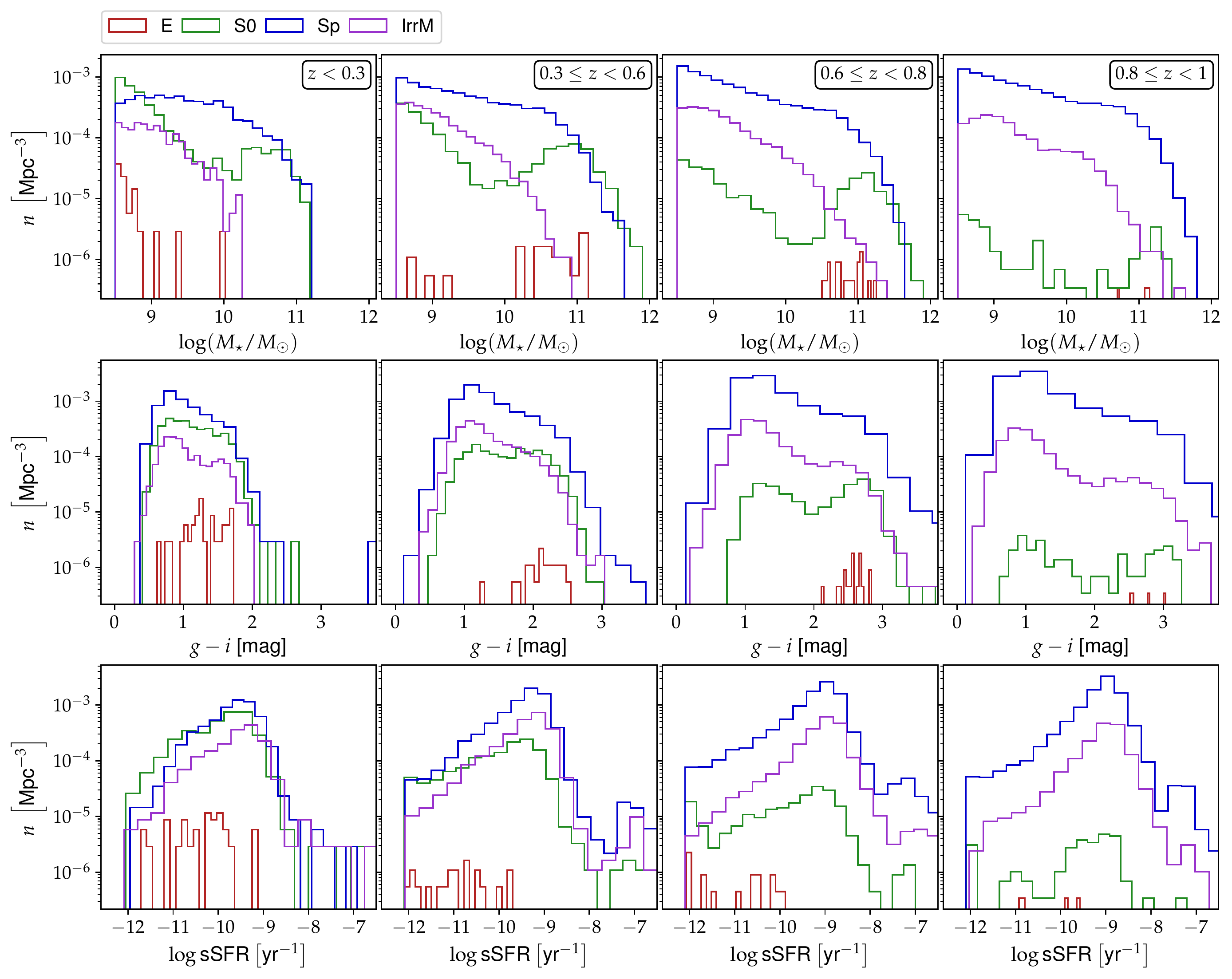}
\caption{{Number density histograms showing the distributions of stellar mass $\log(M_\star/M_\odot)$, colour $g-i$ and sSFR for samples classified as elliptical, S0, spiral and irregular over different redshift ranges.}}
\label{fig:numberdensity}
\end{figure*}

\begin{figure*}
\centering
\includegraphics[scale=0.7]{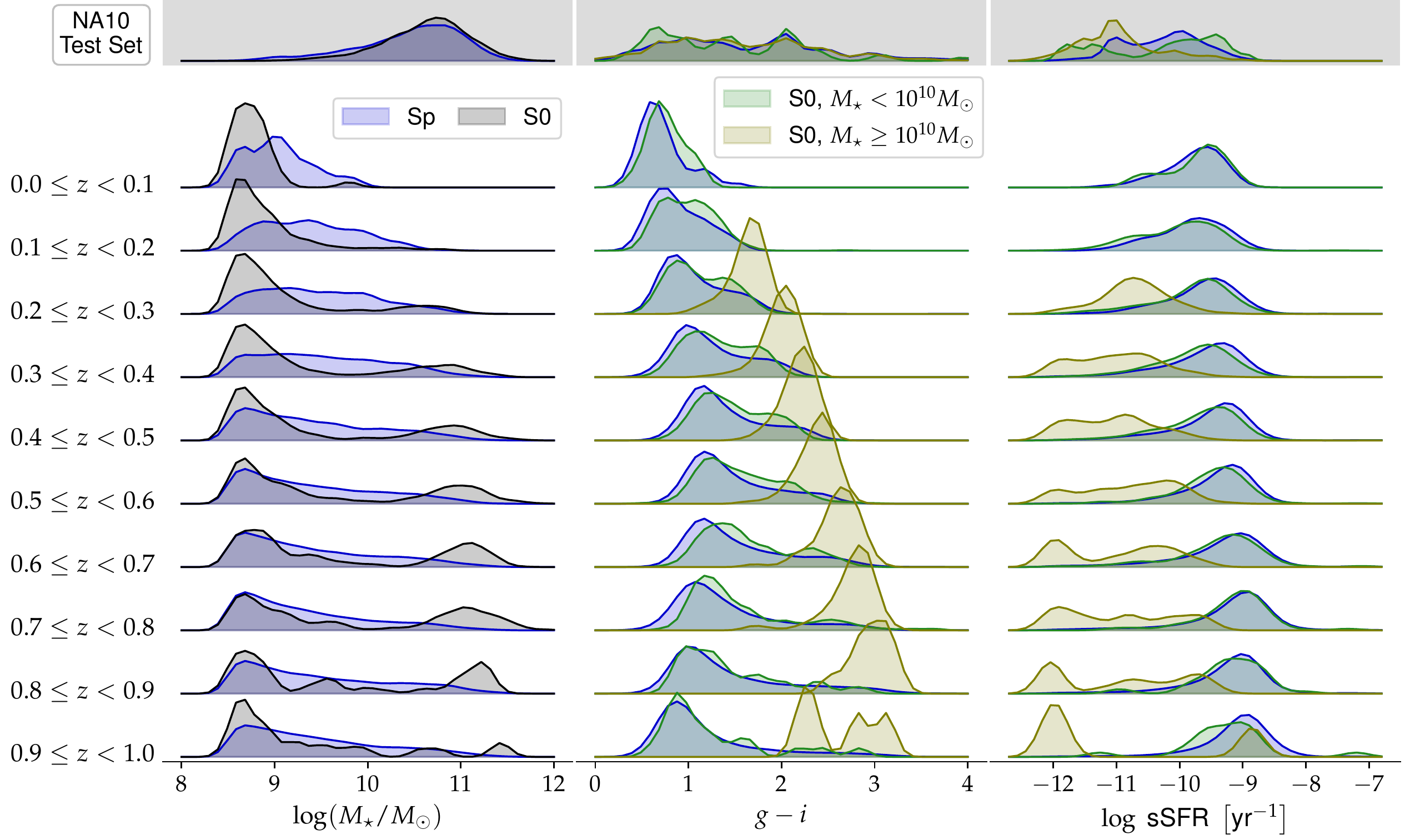}
\caption{Ridgeplots showing the probability density distributions of stellar mass $\log(M_\star/M_\odot)$, colour $g-i$ and sSFR for lenticulars and spirals in finer 0.1$z$ redshift increments from $0 \leq z < 0.1$ to $0.9 \leq z < 1$. The columns for colour and sSFR split the S0s into samples with mass $<10^{10} M_\odot$ in green and $\geq 10^{10} M_\odot$ in olive.}
\label{fig:ridge}
\end{figure*}

Based on our classifications, we are able to determine the fraction of each morphology as a function of redshift for $z < 1$. As described in Section 2.4, these classifications are based on the class corresponding to the highest mean probability across all individual models in the ensemble. Figure \ref{fig:ensemble_fracs} shows the evolution of the morphological fractions with redshift for each morphology, with the shaded regions denoting 1$\sigma$ standard error based on the classification accuracies of Figure \ref{fig:confmat}. For reference at $z \approx 0$, we include the corresponding morphological fractions from the NA10 test set, as classified by our initial CNNs. In both the 3-class and 4-class model, we see that there is a gradual albeit sustained rise in the S0 fraction as redshift decreases, from less than 1\% at $z \approx 1$ to around 40\% at $z \approx 0.1$. The fraction of spirals correspondingly decreases with decreasing redshift. {Both models detect an extremely low number of ellipticals}, while the 4-class model predicts a modest fraction of irregular/miscellaneous samples that peak at around 23\% at $z \approx 0.5$. Compared to the NA10 test set, the COSMOS dataset shows relatively more S0s and much fewer ellipticals at low redshifts. The fraction of spirals at $z \approx 0.1$ also agree well with the overall fraction in NA10; so too does the low fraction of irregulars.

Of the samples with masses $\log(M_\star/M_\odot) > 10$, both models predict the S0 fraction to be slightly lower, and the 4-class model implies a higher fraction of spirals. Approximately 12\% of samples with $\log(M_\star/M_\odot) > 10$ at $z \approx 1$ are classified as irregular/miscellaneous; this fraction all but vanishes as redshift decreases. Consequently, irregulars are mostly below $10^{10} M_\odot$ at low redshifts. Importantly, as shown in Figure \ref{fig:sample-selection}, the total number of samples in COSMOS decreases significantly at low redshifts. This is especially apparent for galaxies above $10^{10} M_\odot$, resulting in much greater uncertainties. There is also a slight dip in the fraction of S0s in the 3-class model that appears to agree well with NA10, however this reduction could also be due to decrease in the number of COSMOS galaxies.

In Figure \ref{fig:ensemble_fracs}, we also saw that the S0 fraction for $\log(M_\star/M_\odot) > 10$ samples began its rise at higher redshifts. We investigate this in more detail in Figure \ref{fig:s0fraction_bymass}, which shows the S0 and spiral fractions for samples in three different mass ranges, as classified by the 4-class ensemble. Figure \ref{fig:s0fraction_bymass} shows that the redshift evolution of the S0 fraction varies significantly for different mass ranges. For high mass samples with $\log(M_\star/M_\odot) \geq 10.5$, the rise in the S0 fraction occurs at higher redshifts, at around $z = 0.9$. For low-mass samples ($\log(M_\star/M_\odot) < 9.5$), the rise is initially more gradual but begins to steadily increase at a similar rate past $z \approx 0.6$. For intermediate mass samples, the rise is much less pronounced, with a final fraction of less than 20\%; less than half of the fraction for the low-mass and high-mass galaxies. These results imply that S0 prevalence depends strongly on stellar mass, favouring both low and high mass galaxies, with a dearth in intermediate galaxies.

Figure \ref{fig:s0fraction_bymass} also shows the redshift evolution of the fraction of spirals. At high redshifts, spirals are dominant across all mass ranges, but as redshift decreases both the low-mass and high-mass spiral fraction fall sharply, while the intermediate-mass spiral fraction remains relatively flat, remaining above 80\% at all redshift ranges. The reduction in the spiral fraction mirrors the increase in the S0 fraction. As with Figure \ref{fig:ensemble_fracs}, Figure \ref{fig:s0fraction_bymass} also gives a $z \approx 0$ reference in terms of the corresponding fractions from the NA10 test set, albeit only for intermediate and high mass ranges due to low numbers of low mass samples. For S0s, the NA10 test set fractions for the intermediate and high-mass ranges are similar, however both have significantly higher uncertainties. Given that there is a dearth of high-mass ($>10^{10.5} M_\odot$) samples for $z < 0.2$ in COSMOS, it's unclear whether the high-mass fraction continues to remain above 40\%, or if it instead decreases. For spirals, the fractions agree well, with a slightly higher fraction of intermediate-mass spirals in COSMOS, although this is only tentative give the greater range of uncertainty at low redshift.

\subsection{Physical Properties}

The results in Figures \ref{fig:ensemble_fracs} and \ref{fig:s0fraction_bymass} indicate that the S0 fraction and its evolution varies with mass. More crucially, these results hint at a potential double-peak mass distribution for S0s. In particular, Figure \ref{fig:s0fraction_bymass} shows that there are more S0s at the low and high mass range, but comparatively fewer S0s at intermediate masses. To investigate this – and see how S0s compare with the other morphologies – Figure \ref{fig:prophist} explores the distribution of masses for {ellipticals, lenticulars, spirals and irregulars}, along with two other markers of morphology: rest frame $g-i$ colour, and specific star formation rates (sSFR), as sourced from the COSMOS2022 catalogue \citep{weaver2022}. Inspecting Figure \ref{fig:prophist}, it can be seen that the mass distribution of S0s indeed exhibits a double peak, with a distinct population of high mass S0s. This second peak is most prominent at higher redshifts and gradually disappears as redshift decreases, likely as a result of the drop in high-mass galaxies in the overall sample. In contrast, the number of low-mass S0s continues to grow in number. This population of high mass S0s appears to correspond to the secondary peaks at redder colours and lower specific star formation rates in the rightmost columns of Figure \ref{fig:prophist}. As the relative number of high-mass S0s decrease with redshift, so too do these peaks disappear. The mass range corresponding to the fewest number of S0s (the ``valley'' of the double-peak distribution) fluctuates slightly about $10^{10} M_\odot$ across the four redshift ranges, but does not vary substantially. Although the mass distributions of each morphology clearly differ for $z < 0.3$, there is only a very slight variation in colour and star formation, with a slightly longer tail of low ssFR S0s. {We note that the drop in the secondary, high-mass peak at lower redshifts may be consequence of COSMOS probing smaller volumes at lower redshifts, hence encountering much fewer high-mass samples (see Figure~\ref{fig:sample-selection}).}

Figure~\ref{fig:prophist} also shows the corresponding physical property distributions in the NA10 test set. The major difference is that the majority of samples in the NA10 test set are high mass samples, while the reverse is true for COSMOS. Even with this in mind, there is little evidence to support the existence of a peak in low-mass S0s in the NA10 test; instead, most S0s have masses above $10^{10} M_\odot$. In NA10, we can see that morphology is well separated by sSFR, but harder to disentangle in terms of $g-i$ colour. That said, the colours and specific star formation rates of NA10 are not directly comparable to that of COSMOS as these are separate surveys, each utilising different instruments, and hence different techniques for measuring such properties.

We can further examine the change in the populations of each morphology by looking at the number density distributions. Figure~\ref{fig:numberdensity} shows the same mass, colour and sSFR distributions as in Figure~\ref{fig:prophist}, albeit in terms of number density. Here we can clearly see the rapid growth in S0s from high redshift to low redshift, as well as a gradual reduction in the number density of spirals. There are an extremely low number of ellipticals present in both Figures~\ref{fig:prophist}~and~\ref{fig:numberdensity}. We note that those that are classified at intermediate redshifts are generally massive, red and quiescents, while at low redshifts there is a significant jump in less massive ellipticals with a wider spread in colour and star formation rates. That said, there are simply too few ellipticals to draw any physical conclusions.

To better illustrate the bimodal distributions of S0 properties, Figure~\ref{fig:ridge} investigates the redshift evolution in the overall frequency density distributions of mass, colour and sSFR in finer redshift bins. {As in Figure~\ref{fig:numberdensity}, we again see that the double-peak mass distribution has mostly declined in prominence by $z \approx 0.3$, likely due to the reduction in massive galaxies in the COSMOS sample at low redshifts.} For colour and sSFR, Figure \ref{fig:ridge} divides S0s into two separate distributions for samples with $M_\star < 10^{10} M_\odot$ (low-mass) and samples with $M_\star \geq 10^{10} M_\odot$ (high-mass). The secondary peak of redder samples in the $g-i$ colour distribution consists almost exclusively of high-mass samples. Low-mass S0s are bluer, but not quite as blue as spirals. Both spirals and S0s have long tails. In the case of star formation, Figure \ref{fig:ridge} suggests that low-mass S0s and spirals are near indistinguishable at low-redshifts, and remain closely coupled throughout all redshift ranges, with S0s exhibiting a slightly longer tail. On the other hand, high mass S0s have lower rates of star formation and are clearly separated from the distribution low-mass S0s. Figure~\ref{fig:ridge} thus shows that the secondary peaks in the colour and sSFR distributions observed in Figure~\ref{fig:prophist} correspond to the high mass samples.

\section{Discussion}

\subsection{Evolution of the S0 Fraction}

Our results support an increase in the S0 fraction with decreasing redshift. In particular, this comes the expense of spiral galaxies, which decrease in prevalence. This result is consistent with previous observational studies in both dense and sparse environments \citep{desai2007, poggianti2009, oesch2010, huertas-company2015}. However, a complete picture that fully describes the formation and evolution of S0s remains to be determined. Studies have demonstrated that there exist multiple mechanisms responsible for S0 formation, such as through a morphological transition from spirals via the faded spiral scenario, \citealt{donofrio2015,rizzo2018,deeley2020}), which is the dominant formation pathway in dense environments and for low-mass S0s. Another formation scenario is that of major merging and/or accretion \citep{querejeta2015,tapia2017,diaz2018,eliche-moral2018}, which is the preferred explanation for the formation of high-mass S0s \citep{fraser-mckelvie2018}. S0s can also form passively through secular evolution and/or bulge growth \citep{kormendy2004,laurikainen2010}, as well as through the passive consumption of gas \citep{rizzo2018}. S0s also exhibit a wide range of physical and dynamical properties \citep{graham2018}, including properties that cannot be explained by faded spirals alone \citep{williams2010,mendez-abreu2018,deeley2021}. As such, the current consensus is that multiple formation pathways are necessary to fully account for the diversity of observed S0s \citep{fraser-mckelvie2018,deeley2020,coccato2022}.

Figures \ref{fig:ensemble_fracs} and \ref{fig:s0fraction_bymass} show that the redshift evolution of the S0 fraction depends strongly on stellar mass. In particular, the fraction of high mass S0s above $10^{10.5} M_\odot$ starts rising at higher $z$ compared to the fraction of low-mass S0s. This suggests that high-mass S0s begin to form earlier than low-mass S0s. The reasons as to why the timescales for S0 formation may vary with stellar mass is unclear. One explanation is that this difference is a consequence of different formation mechanisms. Recent studies have indicated that the formation pathways for S0s may depend on stellar mass \citep{johnston2022}. In particular, \citet{fraser-mckelvie2018, dominguezsanchez2020} find that the formation of high-mass S0s is likely driven by mergers and accretion, while low-mass S0s are more likely to have formed via the faded spiral pathway (see also \citealt{barway2013}). Simulations have demonstrated that low-mass S0s better resemble a spiral progenitor rather than merger remnants \citep{bellstedt2017}.

\citet{mendez-abreu2018} studied the evolution of high-mass S0s and suggested that their formation occurs at earlier epochs, and are largely driven by high-redshift dissipational processes \citep{johnston2020}. The study by \citet{oesch2010}, which examined S0s in COSMOS, determined that secular evolution is also an important driver of morphological transitions away from spirals into S0s, especially for low-mass galaxies. However, we note that formation pathways are also strongly dependent on environment \citep{mishra2019}, with the faded spiral pathway being especially dominant in dense environments such as clusters \citep{johnston2014,coccato2022}. Further studies are needed to ascertain the full extent of passive disc fading as a formation mechanism for S0s in low-density environments.

Figure \ref{fig:s0fraction_bymass} also showed that the S0 fraction in intermediate mass ($10^{9.5} M_\odot$ to $10^{10} M_\odot$) galaxies is not only the smallest, but also rises with the slowest rate. Furthermore, Figure \ref{fig:s0fraction_bymass} also illustrates that these intermediate-mass samples are overwhelmingly spiral galaxies. These results suggest that the formation pathways for S0s are more effective for low and high-mass galaxies.

\subsection{A Bimodal Population}

\begin{figure}
\centering
\includegraphics[scale=0.535]{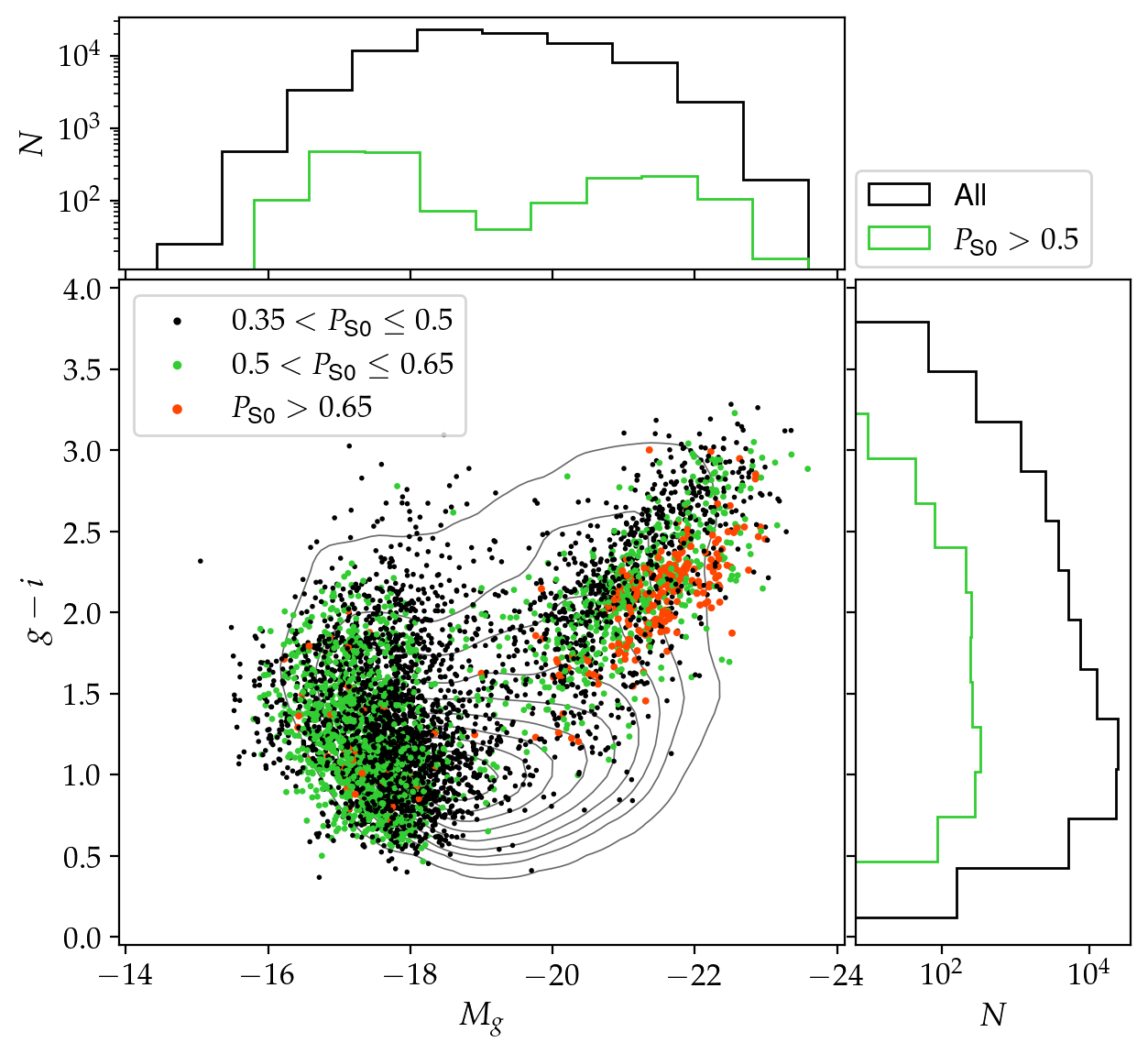}\\
\includegraphics[scale=0.525]{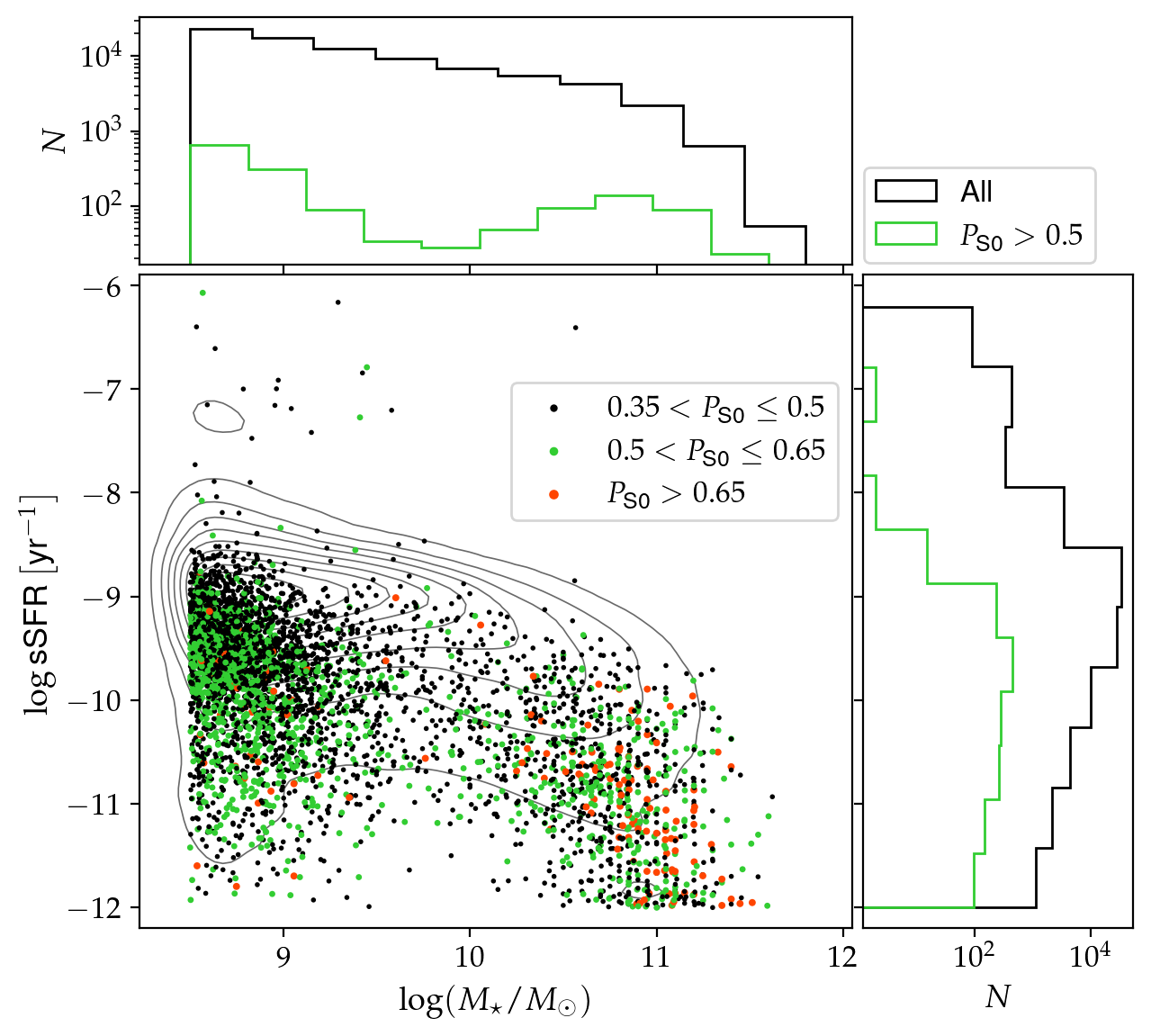}
\caption{(Top) Absolute $g$ magnitude $M_g$ vs. colour $g-i$. (Bottom) Stellar mass $\log(M_\star/M_\odot)$ vs. sSFR. Contours denote smoothed kernel density estimates of the distribution of the entire sample. {Samples with $P_{\text{S0}} > 0.35$ are directly plotted and shaded according to their mean S0 output probability, as predicted by the augmented 4-class model ensemble.}}
\label{fig:cmd-mssfr}
\end{figure}

\begin{figure}
\centering
\includegraphics[scale=0.57]{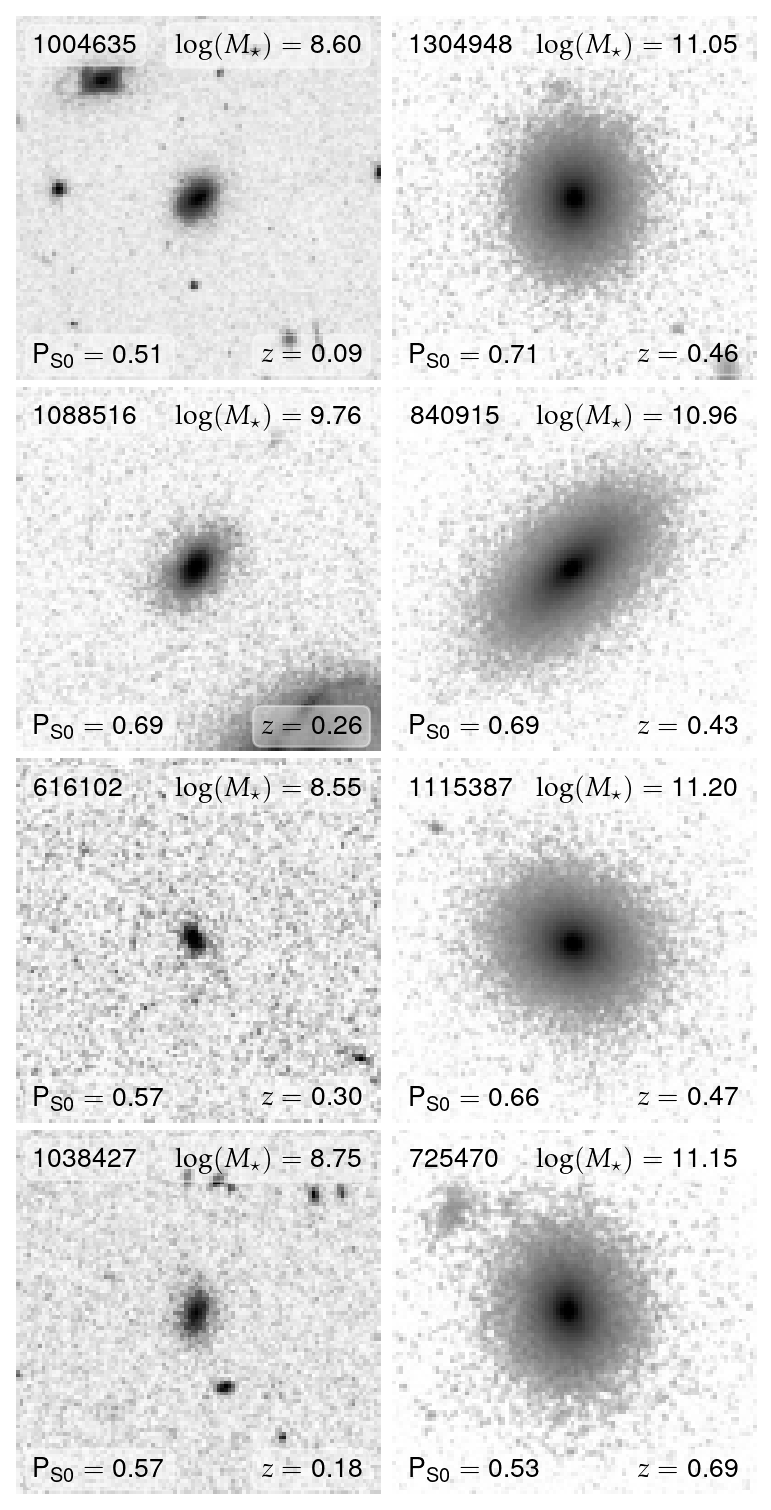}
\caption{Random selection of example S0 galaxies classified with at least a confidence $P_{\text{S0}} > 0.5$. The left column shows a selection of low mass S0s, and the right row shows a selection of high mass S0s. Samples are annotated with their COSMOS ID in the top left, stellar mass in the top right in units of $\log(M_\star/M_\odot)$, confidence $P_{\text{S0}}$ in the bottom left, and redshift $z$ in the bottom right.}
\label{fig:exlowhighs0s}
\end{figure}

Based on our classifications and the results of Figures \ref{fig:prophist} and \ref{fig:ridge}, we have unearthed tentative evidence supporting the existence of two distinct populations of S0s, distinguished by their bimodal mass distributions, and subsequent differences for both colour and specific star formation rate. Figure \ref{fig:cmd-mssfr} illustrates these properties with both a galaxy colour-magnitude diagram and a plot of sSFR versus stellar mass. From the colour-magnitude diagram we can see two distinct populations of high-confidence S0 galaxies: a population of less luminous, bluer S0s; and a population of more luminous, redder S0s, with $M_g \approx 19$ serving as an approximate divider separating the two populations. While there do exist S0s in the sparse transition region between these populations ($-18 \lesssim M_g \lesssim -20$), {these S0s tend to have much lower classification confidences ($P_{\text{S0}} < 0.5$) compared to those in the two hotspots. This transition region is instead largely dominated by spirals. The galaxies with the highest classification confidences $P_{\text{S0}} > 0.65$ are almost all concentrated in the redder, higher-luminosity population.} Furthermore, we note that the low-luminosity samples have a greater spread in $g-i$ colour compared to the high-luminosity samples, which are instead predominantly red ($g-i \gtrsim 2$).

Figure \ref{fig:cmd-mssfr} also illustrates sSFR, from which we can see that the majority of S0s have lower rates of star formation, and are mostly found below the galaxy main sequence. However, we also see evidence for star-forming S0s with star formation rates in line with spirals, as suggested by the frequency densities in Figure \ref{fig:ridge}. In particular, these star-forming S0s are mostly low-mass S0s. There is a greater range of sSFR values for low-mass S0s compared to high-mass S0s, the latter of which are mostly quiescent, which is consistent with a merger-driven formation pathway for high-mass S0s. From Figure \ref{fig:cmd-mssfr}, we infer that the low-mass population includes both star-forming and passive S0s. This implies that there are multiple formation pathways for low-mass S0s.

We also note that there is a region of sparse sampling between $9 < \log(M_\star / M_\odot) < 10.5$ in which there are comparatively fewer galaxies. This region coincides with the valley in the S0 mass distribution at intermediate masses. Indeed, from Figure \ref{fig:s0fraction_bymass}, we have established that the majority of these intermediate mass galaxies are spirals, which is concordant with the lack of quiescent galaxies in this mass range, as seen in Figure \ref{fig:cmd-mssfr}. This explains the low S0 fraction for intermediate mass samples.

Figure \ref{fig:exlowhighs0s} shows a random selection of low-mass and high-mass S0 galaxies, as sampled from the two subpopulations of high-mass quiescent S0s, and low-mass star-forming S0s. The high-mass examples are share the typical characteristics of an S0 galaxy; featureless, well-defined disc components with strong central bulges. By comparison, the low-mass examples are considerably more compact. Some such as 1088516 resemble dwarf lenticular galaxies, while others such as 616102 are harder to visually discern, with comparatively lower classification accuracies.

Recent work by \citet{tous2020}, based on machine learning classifications, also uncovered two subpopulations of S0s with distinct properties, some of which exhibited star formation rates more in-line with late type spirals. \citet{tous2020} conclude that star-forming S0s may not be as rare as first thought. Indeed, previous studies have searched for star-forming early types and concluded that they are rare, but most abundant at low stellar masses $M_\star < 10^9 M_\odot$ \citep{kannappan2009, schawinski2009}. Unlike passive S0s, star forming S0s cannot be readily explained by the faded spiral formation mechanism. It is believed that the star formation in such low-mass early types is consistent with starbursts triggered by minor merging \citep{wei2010} or accretion from gas-rich companions \citep{marino2011}. Recent studies examining star-forming S0s (e.g. \citealt{rathore2022}) have suggested that they could be quenched objects currently undergoing renewed star formation. The majority of star-forming S0s analysed by \citet{rathore2022} had masses less than $10^{10.25} M_\odot$. \citet{dominguezsanchez2020} also found a bimodality in the S0 population, with high-mass galaxies exhibiting strong velocity dispersions with little to no metallicity gradients, while low-mass galaxies had flat velocity dispersion profiles and strong metallicity gradients. {\citet{cassata2007} also found evidence for an anomalous population of ``blue spheroids'', which may be consistent with a mixture of spirals, dwarf ellipticals and S0s galaxies subject to strong bursts of star formation.}

\subsection{Our Deep Learning Approach}

\subsubsection{NA10 and COSMOS}

\begin{figure}
\centering
\includegraphics[scale=0.53]{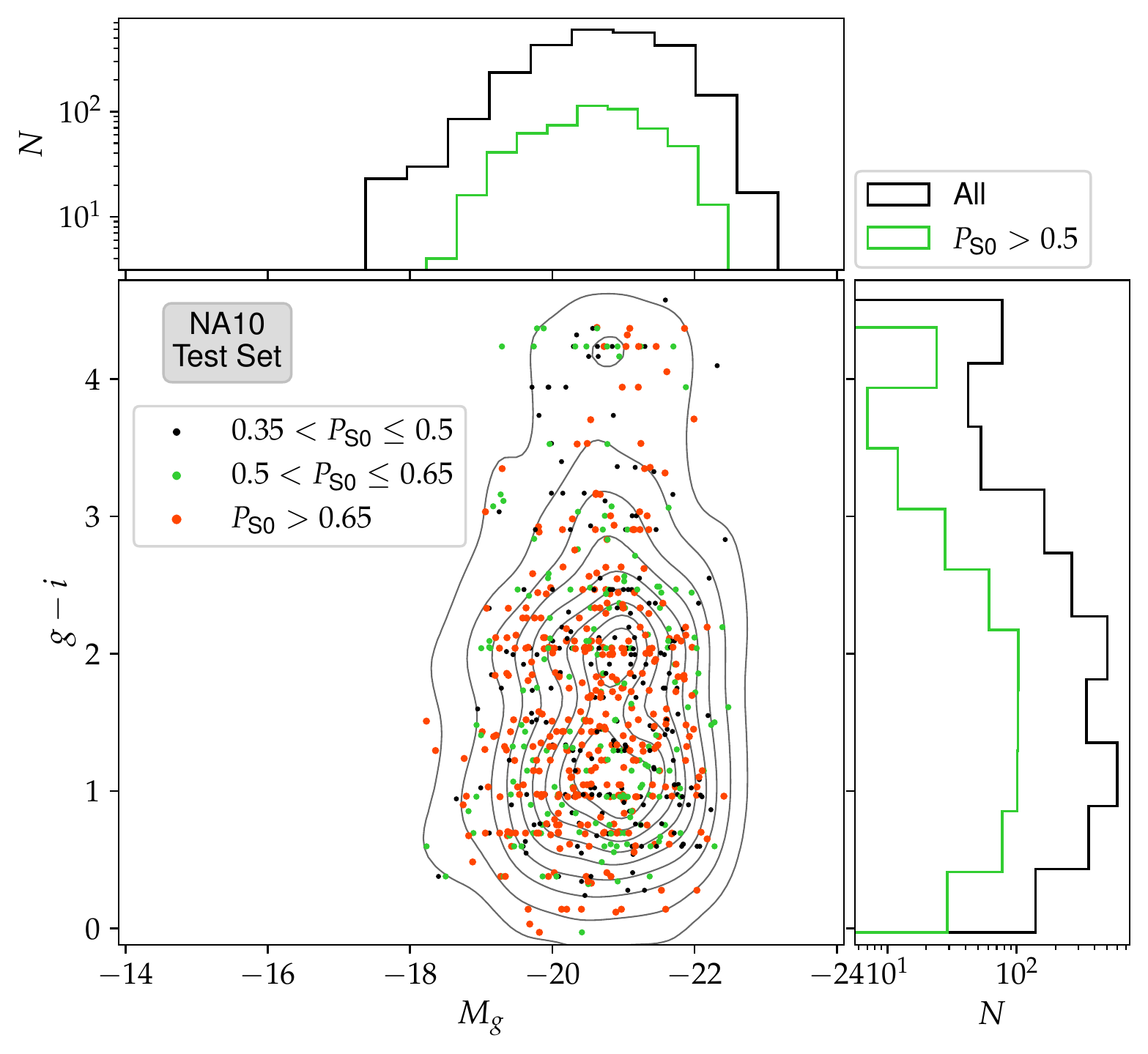}\\
\includegraphics[scale=0.525]{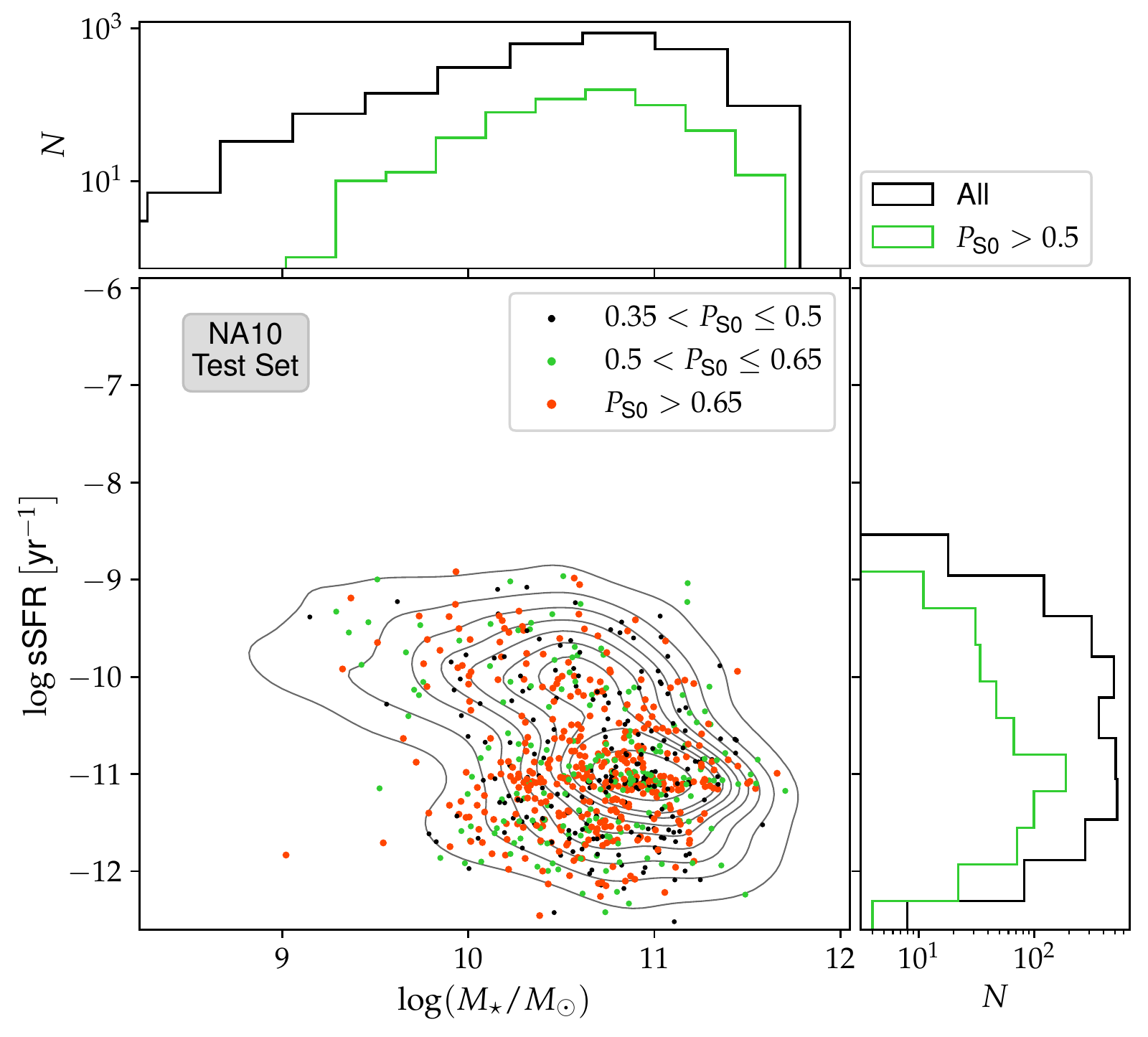}
\caption{Absolute $g$ magnitude $M_g$ vs. colour $g-i$ (top) and stellar mass $\log(M_\star/M_\odot)$ vs. sSFR (bottom) for galaxies in the NA10 test set. {As in Figure \ref{fig:cmd-mssfr}, the contours are drawn with respect to the full sample. Galaxies with $P_{\text{S0}} > 0.35$ are directly plotted and shaded according to their classification confidence, as predicted by the initial 4-class base model.}}
\label{fig:cmd-na10}
\end{figure}

The deep learning classifiers used to classify the images from our COSMOS sample were adapted from base models that were initially trained on SDSS $g$-band images with known morphologies from the NA10 dataset. As a baseline to aid in interpreting our results so far, Figure \ref{fig:cmd-na10} shows the same diagrams as in Figure \ref{fig:cmd-mssfr}, but this time for samples from the NA10 test set, as classified by the base models. We caution that the values, especially sSFRs, are not directly comparable due to differences in the methods used to obtain them. Furthermore, NA10 samples a much higher mass range compared to our COSMOS sample. That said, Figure \ref{fig:cmd-na10} offers an insight into the original capabilities of the model. In particular, even despite the poor sampling at lower masses, there is no indication of a secondary peak in mass. Most S0s are quiescent and clustered between $10^{10}$ and $10^{11} M_\odot$, with a peak around $10^{10.5} M_\odot$, however there are also several S0s with star formation rates in the main sequence, including samples with masses less than $10^{9.5} M_\odot$. However, there are too few low-mass samples in NA10 test set – and the entire NA10 sample as a whole – to clearly establish any bimodal mass distribution, nor the existence of low-mass, star forming S0s. On the other hand, the COSMOS sample is awash with low-mass samples. The fact that these models are able to identify S0s among these low-mass samples, despite there being so few in NA10, is significant, for if the models were simply trained to mimic NA10 then Figure \ref{fig:cmd-mssfr} would look markedly different.

{Both our models detected an extremely low number of ellipticals, to the point where we could not reliably establish either their properties or evolution. This is a significant issue given that ellipticals are plentiful in NA10, as shown in Figures~\ref{fig:prophist}. There are several factors that may account for this, not least the inherent differences between the NA10 dataset and low-redshift COSMOS data. Importantly, the NA10 dataset is not volume corrected and hence the fractions of ellipticals are not representative. We note that studies examining volume corrected SDSS galaxies have also obtained low fractions for ellipticals \citep{wilman2012}. Furthermore, the NA10 ellipticals are predominantly massive with stellar masses $\log(M_\star/M_\odot) > 10.5$, however COSMOS samples very few massive galaxies at low redshifts. Previous studies have also shown that the fraction of early types is considerably greater in dense environments such as clusters rather than in the field \citep{fasano2000, wilman2009, poggianti2009, kovac2010}, with \citep{desai2007} finding little evolution in the fraction of ellipticals at low redshifts. In the COSMOS field environment, ellipticals should constitute at least 15\% of the total number of galaxies \citep{oesch2010}, increasing significantly for massive galaxies. We ought to be detectable at higher redshifts with our deep learning models, however this is where noise likely has a substantial impact.}

\subsubsection{The Impact of Noise on Classification Uncertainty}

\begin{figure*}
\centering
\includegraphics[scale=0.6]{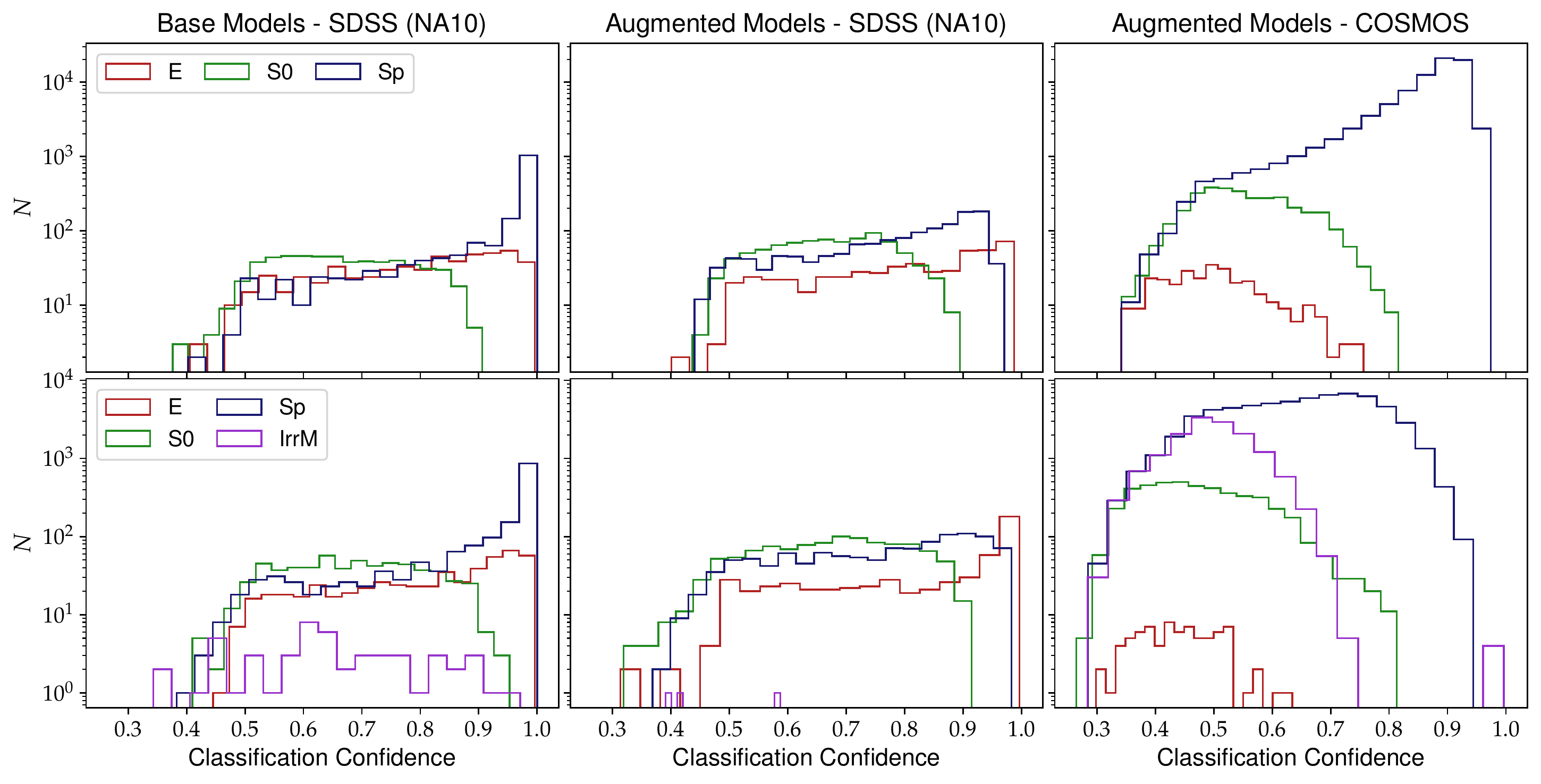}
\caption{Distribution of the classification confidences of all predictions for each morphological class in the 3-class (top row) and 4-class (bottom row) models. The left column shows the initial, base models as evaluated on the original images from the NA10 test set. The middle column shows the results of the noise-augmented models, as evaluated on noisy images (also from the NA10 test set). The right column shows the noise augmented models as applied to the entire COSMOS dataset.}
\label{fig:conf34}
\end{figure*}
\begin{figure}
\centering
\includegraphics[scale=0.47]{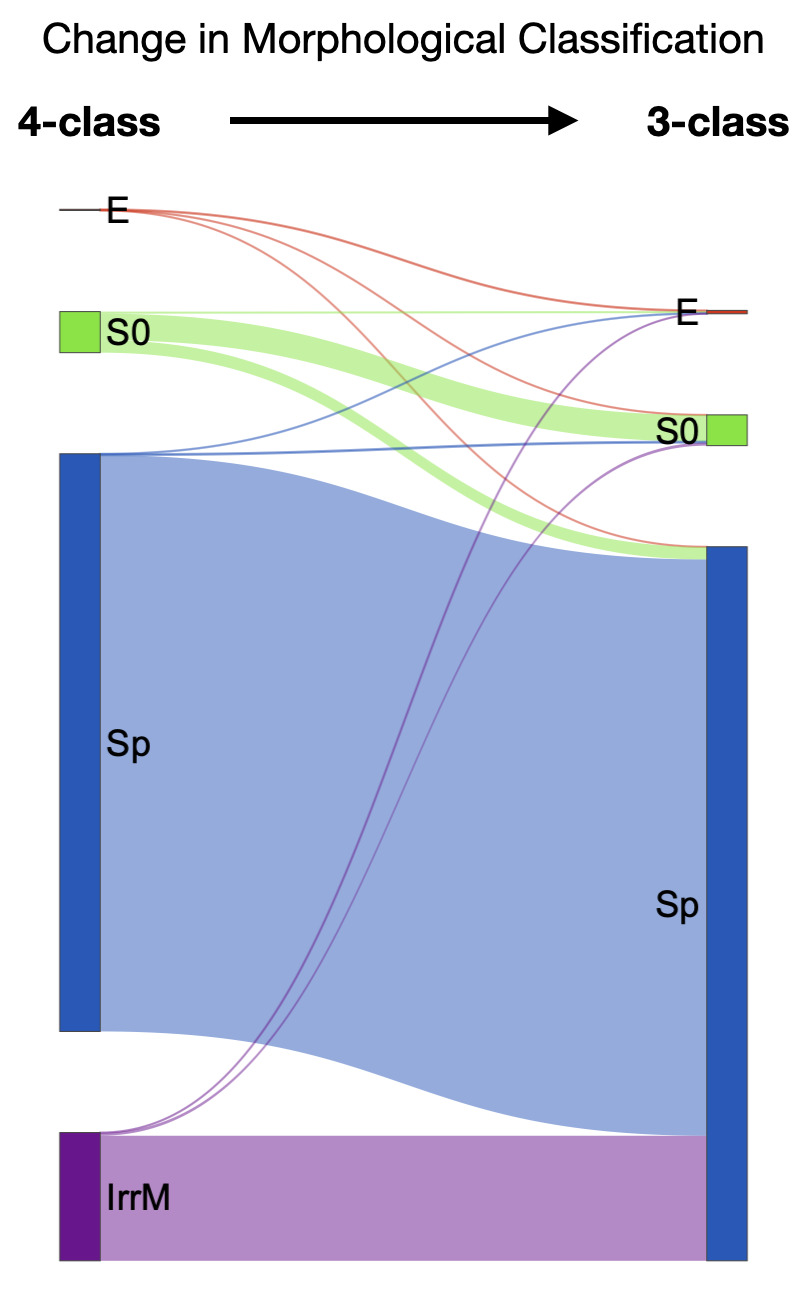}
\caption{{Sankey diagram showing the change in classifications for all galaxies when predicted by the 4-class model, to that predicted by the 3-class model. Vertical bands denote morphological categories, and lines denote flow. Both are sized linearly in proportion to the number of galaxies.}}
\label{fig:sankey}
\end{figure}

As aforementioned in Section 3, the data augmentation workflow is designed to add a random level of Gaussian noise to each image for each batch in each training epoch, such that the images collectively resemble the noise levels of the images in the COSMOS dataset {(see Appendix A for further details on the exact procedure)}. As was observed in Figure \ref{fig:confmat}, we found that the S0 accuracy increased, while the accuracy of the other classes decreased. In our 4-class model, irregulars were almost all misclassified as either S0 or spiral, and up to 30\% of true spirals were misclassified as S0.

We can examine the impact of the noise augmentation on a macro level by considering the distribution of classification confidences for each predicted class. Figure~\ref{fig:conf34} shows the distribution of classification confidences for all samples classified into each morphological class, i.e. $P_{\text{E}}$ for all samples classified as E, $P_{\text{S0}}$ for all samples classified as S0, etc. {A direct comparison can be made by examining the application of the base model on the NA10 test set, as well as the noise-augmented models on noisy NA10 images; this is shown in the first two columns of Figure~\ref{fig:conf34}. Here, noisy NA10 images simply refers to images that underwent the same artificial noise injection as that used to train the noise-augmented models (see Appendix A).} The most immediate difference is that irregulars all but vanish for the 4-class noise augmented model. In the base models, we see a very strong peak for spirals with confidences of around 1; such a peak of near-certain predictions is indicative of overfitting. The distributions for spirals are comparatively smoother and flatter for the noise augmented models, and the maximum confidence for spirals is less than 1. This is likely directly due to the noise impacting on the ability to discern spirals. We also see that S0s cannot be classified with a confidence greater than 0.9. Ellipticals maintain roughly the same distribution, with a slight peak in higher confidences for the 4-class noise augmented model. While most samples are classified with confidences greater than 0.5, there is a tail of samples with confidences below this, indicating that there are images in the NA10 test set that are difficult for the model to classify.

The last column in Figure~\ref{fig:conf34} shows the distribution of prediction confidences for all 85,378 images in the COSMOS dataset. We can conclude that the S0s in COSMOS are harder to classify confidently; no samples have confidences beyond 0.8. Although the classification accuracies of the noise augmented models is higher for S0s, this comes at the cost of higher uncertainty, which is no surprise given the fractions of spirals and ellipticals misclassified as S0s as seen in Figure \ref{fig:confmat}. {The ellipticals in COSMOS are classified with the lowest confidences, with $P_{\text{E}} < 0.65$ in the 4-class model. This is despite ellipticals being relatively easy to classify in NA10, even with the augmented models (see middle column of Figure~\ref{fig:conf34}). This suggests that either COSMOS contains a critical shortage of elliptical galaxies, or that ellipticals are much harder to accurately detect, even with a model trained specifically to be robust with respect to noise (especially if one accepts, as visual criteria for classification, that ellipticals are featureless spheroids). Furthermore, although our base models detect very few irregulars in the SDSS images, our augmented models are able to detect far more irregulars in COSMOS. This is likely a combination of factors, including that of increased noise paired with more limited image qualities and resolutions, however it is also possible that there are simply more inherently irregular galaxies in COSMOS.}

It is also worth examining the degree to which the overall classifications differ between the 3-class and 4-class models. We saw individual examples of this in Figure \ref{fig:examples_differ}. To illustrate the overall impact, Figure \ref{fig:sankey} shows the overall changes in classifications for galaxies in COSMOS initially classified as E, S0, Sp and IrrM by the 4-class model. The majority of classifications are unchanged, however there is a sizeable fraction of S0s that the 3-class model classifies as spirals. A degree of misclassification for S0s is expected given the confusion matrices in Figure \ref{fig:confmat}, yet this is likely also exacerbated given the dominance of spirals in the 3-class model classifications. Importantly, the galaxies with different classifications were found to have substantially lower confidences (mean $P_{\text{S0}} \approx 0.4$, both models) than those whose classifications remained unchanged ($P_{\text{S0}} \approx 0.6$).

This study focuses exclusively on dealing with different noise levels since this has a substantial negative impact on CNN accuracy, as demonstrated Figure \ref{fig:noiseacc}. However, differing noise levels is not the only consideration when adapting a model to classify images from different datasets. Other factors include the impacts of different angular and/or physical resolutions, as well as classifying images at different wavelengths, either due to different redshifts ranges or different observational bands. Noise is a destructive process that directly impacts on the CNN's ability to extract meaningful features, particularly in the first convolutional layer. Differing resolutions due to different PSFs can be easily characterised by a convolutional kernel, whereas noise cannot. Although the effects of different resolutions and wavelengths cannot be completely ignored, we argue that the impacts of these for our NA10 SDSS to COSMOS transfer learning is limited. This is due to the low input resolution of the CNN of $100 \times 100$ pixels, which is smaller than the average image pixel size of our COSMOS galaxies prior to data preprocessing ($153 \times 153$ pixels). Hence, for an average image, the PSF is on the order of approximately 0.1'' \citep{koekemoer2007}, or around 3 pixels. This is further reduced to around 2 pixels when the images are downsized, which is well below the size of the first layer's convolutional kernels (see Figure \ref{fig:cnn-schematic}). For the original SDSS training images, the median PSF FWHM is around 1.3'' \citep{abazajian2009} corresponding to approximately 0.78kpc at $z = 0.03$, or less than 2 pixels given our fixed 0.5 kpc physical pixel scale. The majority of the SDSS images were likewise downsized as part of the preprocessing. Furthermore, our model is restricted to single-band images, and we only classify samples up to an upper redshift limit of $z = 1$. This limits the impacts of k-correction, which become increasingly significant for very high values of $z$.

{It is difficult to determine the impact of these differing scales on our results, however these factors cannot be ruled in adversely impacting our model's performance, as demonstrated with the reduction in overall classification confidences in Figure~\ref{fig:conf34}. We argue that noise has the most dominant impact, and its effects become progressively worse at higher and higher redshifts, where resolution is increasingly limited.} Taken together, the impacts of noise, resolution and k-correction will ultimately place upper limits on the maximum redshift to which a deep learning approach based on images is suitable. Future studies utilising high-resolution JWST imaging will likely go beyond such existing limits, and enable more accurate deep learning classifications for very high-redshift galaxies \citep{ferreira2022,kartaltepe2022,robertson2022}.

\section{Conclusion}

In this work, we investigated the redshift evolution of the S0 fraction for $z < 1$ in COSMOS, through the application of convolutional neural networks to classify images of galaxies. Our transfer learning approach, as facilitated by data augmentation, enables us to adapt existing models trained on low-redshift SDSS images to classify high-redshift COSMOS images, all the while leveraging our existing images and labels of known morphologies. While previous studies have demonstrated it is possible to use transfer learning to adapt models to classify low-redshift images from different surveys \citep{dominguezsanchez2019}, ours is the first to show it is possible to adapt models to classify images across different redshift regimes. Our key findings are summarised as follows.

\begin{enumerate}
\item We have found that there is a sustained rise in the overall S0 fraction from $z = 1$ to $z = 0.1$ from less than 1\% to around 40\%. Furthermore, this rise comes at the expense of spiral galaxies, for which the fraction instead decreases.\medskip

\item The onset of the growth in the S0 fraction occurs at higher redshifts for high-mass ($\geq10^{10.5} M_\odot$) samples at around $z = 0.9$, rising sharply to 45\% by $z = 0.4$ where it remains stable. For low-mass samples ($<10^{9.5} M_\odot$), the rise in the S0 fraction is delayed, reaching a comparable level by $z = 0.2$. Between $z=0.5$ and $z=0.7$, there are nearly twice as many high-mass S0s as there are low-mass S0s. We conclude that, in general, high-mass S0s evolved earlier than low-mass S0s. Previous studies have established that S0 formation pathways depend strongly on mass, and so it is likely that this difference is as a result of varying timescales for different S0 transformation processes.\medskip

\item At low redshifts, the S0 fraction in intermediate mass S0s ($10^{9.5} M_\odot$ to $10^{10.5}M_\odot$) is roughly half that of the fraction in high and low-mass samples. Overall, the S0 fraction is highest in the low mass and high mass range. On the other hand, the fraction of spirals is highest for intermediate masses (80\%), while lower for low and high masses (around 50\% and 55\% respectively).\medskip

\item We have found a bimodal mass distribution in our classified S0 galaxies, such that that they constitute two largely distinct populations: high-mass S0s that are almost all red and quiescent; and low-mass S0s that are generally bluer with higher star formation. We have found that low-mass S0 population includes both passive S0s and star-forming S0s, the latter of which cannot be solely explained by the faded spiral mechanism. As such, we suggest that there are a range of physical processes responsible for the formation and evolution of low-mass S0s, including processes that do not completely quench star formation.\medskip

\item We have demonstrated the effectiveness of transfer learning and data augmentation in adapting our initial models, pretrained on SDSS images of NA10 samples, to classify COSMOS images, which have significantly higher levels of noise. Through fine-tuning, we trained a new ensemble of models with artificially noisy SDSS images such that they replicate the characteristics of the COSMOS images. Importantly, this method allows us to utilise the known NA10 morphologies for training. We have found that this method led to a 10\% increase in S0 accuracy on the NA10 test set. Even at moderate levels of noise, our augmented models dramatically outperform the initial base models, with overall classification accuracies around 70 to 80\%, compared to less than 30\% for the base models without noise augmentation. We thus conclude that transfer learning is crucial for CNNs to classify images from different surveys where there is a major difference in noise levels.

\end{enumerate}

\section*{Acknowledgements}

We wish to acknowledge and thank the anonymous referee for their constructive feedback that helped to improve the presentation of the paper. MKC acknowledges the support of the Australian Government Research Training Programme at the University of Western Australia. Our models were developed with \textsc{TensorFlow} \citep{abadi2016} and \textsc{Keras} \citep{chollet2015}, and trained on an Nvidia RTX 3000 series GPU. Furthermore, significant parts of this work utilised the following Python packages: \textsc{Optuna} \citep{akiba2019}, \textsc{Pillow} \citep{vankemenade2022}, \textsc{Numpy} \citep{harris2020}, \textsc{Matplotlib} \citep{hunter2007}, and \textsc{Astropy} \citep{theastropycollaboration2013}. We further acknowledge COSMOS, the publicly available COSMOS2020 catalogue \citep{weaver2022}, and the HST-ACS image data \citep{koekemoer2007}.

\section*{Data Availability}

This research utilised HST-ACS imaging \citep{koekemoer2007} and catalogues from the Cosmological Evolution Survey \citep{laigle2016,weaver2022}, which are publicly accessible on the NASA/IPAC Infrared Science Archive
\url{https://irsa.ipac.caltech.edu/data/COSMOS/overview.html}.
The COSMOS2020 catalogue is accessible at \url{https://cosmos2020.calet.org/}. The catalogue of morphologies used to train the initial model is described in \citet{nair2010}. Specific data pertinent to this work will be made available upon reasonable request to the author.



\bibliographystyle{mnras}
\bibliography{mybib_final} 

\begin{thebibliography}{}
\makeatletter
\relax
\def\mn@urlcharsother{\let\do\@makeother \do\$\do\&\do\#\do\^\do\_\do\%\do\~}
\def\mn@doi{\begingroup\mn@urlcharsother \@ifnextchar [ {\mn@doi@}
  {\mn@doi@[]}}
\def\mn@doi@[#1]#2{\def\@tempa{#1}\ifx\@tempa\@empty \href
  {http://dx.doi.org/#2} {doi:#2}\else \href {http://dx.doi.org/#2} {#1}\fi
  \endgroup}
\def\mn@eprint#1#2{\mn@eprint@#1:#2::\@nil}
\def\mn@eprint@arXiv#1{\href {http://arxiv.org/abs/#1} {{\tt arXiv:#1}}}
\def\mn@eprint@dblp#1{\href {http://dblp.uni-trier.de/rec/bibtex/#1.xml}
  {dblp:#1}}
\def\mn@eprint@#1:#2:#3:#4\@nil{\def\@tempa {#1}\def\@tempb {#2}\def\@tempc
  {#3}\ifx \@tempc \@empty \let \@tempc \@tempb \let \@tempb \@tempa \fi \ifx
  \@tempb \@empty \def\@tempb {arXiv}\fi \@ifundefined
  {mn@eprint@\@tempb}{\@tempb:\@tempc}{\expandafter \expandafter \csname
  mn@eprint@\@tempb\endcsname \expandafter{\@tempc}}}

\bibitem[\protect\citeauthoryear{Abadi et~al.,}{Abadi et~al.}{2016}]{abadi2016}
Abadi M.,  et~al., 2016, preprint
  (\href{https://arxiv.org/abs/1603.04467}{arXiv:1603.04467})

\bibitem[\protect\citeauthoryear{Abazajian et~al.,}{Abazajian
  et~al.}{2009}]{abazajian2009}
Abazajian K.~N.,  et~al., 2009, \mn@doi [ApJS] {10.1088/0067-0049/182/2/543},
  182, 543

\bibitem[\protect\citeauthoryear{Ackermann, Schawinski, Zhang, Weigel  \&
  Turp}{Ackermann et~al.}{2018}]{ackermann2018}
Ackermann S.,  Schawinski K.,  Zhang C.,  Weigel A.~K.,   Turp M.~D.,  2018,
  \mn@doi [MNRAS] {10.1093/mnras/sty1398}, 479, 415

\bibitem[\protect\citeauthoryear{Adelman-McCarthy et~al.,}{Adelman-McCarthy
  et~al.}{2006}]{adelman-mccarthy2006}
Adelman-McCarthy J.~K.,  et~al., 2006, \mn@doi [ApJSS] {10.1086/497917}, 162,
  38

\bibitem[\protect\citeauthoryear{Akiba, Sano, Yanase, Ohta  \& Koyama}{Akiba
  et~al.}{2019}]{akiba2019}
Akiba T.,  Sano S.,  Yanase T.,  Ohta T.,   Koyama M.,  2019, in Proceedings of
  the 25th {{ACM SIGKDD International Conference}} on {{Knowledge Discovery}}
  \& {{Data Mining}}. {ACM}, {Anchorage AK USA}, pp 2623--2631,
  \mn@doi{10.1145/3292500.3330701}

\bibitem[\protect\citeauthoryear{Arnouts \& Ilbert}{Arnouts \&
  Ilbert}{2011}]{arnouts2011}
Arnouts S.,  Ilbert O.,  2011, Astrophysics Source Code Library, p.
  ascl:1108.009

\bibitem[\protect\citeauthoryear{Bait, Barway  \& Wadadekar}{Bait
  et~al.}{2017}]{bait2017}
Bait O.,  Barway S.,   Wadadekar Y.,  2017, \mn@doi [MNRAS]
  {10.1093/mnras/stx1688}, 471, 2687

\bibitem[\protect\citeauthoryear{Barr, Bedregal, {Arag{\'o}n-Salamanca},
  Merrifield  \& Bamford}{Barr et~al.}{2007}]{barr2007}
Barr J.~M.,  Bedregal A.~G.,  {Arag{\'o}n-Salamanca} A.,  Merrifield M.~R.,
  Bamford S.~P.,  2007, \mn@doi [A\&A] {10.1051/0004-6361:20077151}, 470, 173

\bibitem[\protect\citeauthoryear{Barway, Wadadekar, Vaghmare  \&
  Kembhavi}{Barway et~al.}{2013}]{barway2013}
Barway S.,  Wadadekar Y.,  Vaghmare K.,   Kembhavi A.~K.,  2013, \mn@doi
  [MNRAS] {10.1093/mnras/stt478}, 432, 430

\bibitem[\protect\citeauthoryear{Beck et~al.,}{Beck et~al.}{2018}]{beck2018}
Beck M.~R.,  et~al., 2018, \mn@doi [MNRAS] {10.1093/mnras/sty503}, 476, 5516

\bibitem[\protect\citeauthoryear{Bekki}{Bekki}{1998}]{bekki1998}
Bekki K.,  1998, \mn@doi [ApJ] {10.1086/311508}, 502, L133

\bibitem[\protect\citeauthoryear{Bekki \& Couch}{Bekki \&
  Couch}{2011}]{bekki2011}
Bekki K.,  Couch W.~J.,  2011, \mn@doi [MNRAS]
  {10.1111/j.1365-2966.2011.18821.x}, 415, 1783

\bibitem[\protect\citeauthoryear{Bekki, Couch  \& Shioya}{Bekki
  et~al.}{2002}]{bekki2002}
Bekki K.,  Couch W.~J.,   Shioya Y.,  2002, \mn@doi [ApJ] {10.1086/342221},
  577, 651

\bibitem[\protect\citeauthoryear{Bellstedt, Forbes, Foster, Romanowsky, Brodie,
  Pastorello, Alabi  \& Villaume}{Bellstedt et~al.}{2017}]{bellstedt2017}
Bellstedt S.,  Forbes D.~A.,  Foster C.,  Romanowsky A.~J.,  Brodie J.~P.,
  Pastorello N.,  Alabi A.,   Villaume A.,  2017, \mn@doi [MNRAS]
  {10.1093/mnras/stx418}, 467, 4540

\bibitem[\protect\citeauthoryear{Borlaff et~al.,}{Borlaff
  et~al.}{2014}]{borlaff2014}
Borlaff A.,  et~al., 2014, \mn@doi [A\&A] {10.1051/0004-6361/201424299}, 570,
  A103

\bibitem[\protect\citeauthoryear{Bundy, Ellis  \& Conselice}{Bundy
  et~al.}{2005}]{bundy2005}
Bundy K.,  Ellis R.~S.,   Conselice C.~J.,  2005, \mn@doi [ApJ]
  {10.1086/429549}, 625, 621

\bibitem[\protect\citeauthoryear{Cappellari et~al.,}{Cappellari
  et~al.}{2011}]{cappellari2011}
Cappellari M.,  et~al., 2011, \mn@doi [MNRAS]
  {10.1111/j.1365-2966.2011.18600.x}, 416, 1680

\bibitem[\protect\citeauthoryear{Cassata et~al.,}{Cassata
  et~al.}{2007}]{cassata2007}
Cassata P.,  et~al., 2007, \mn@doi [ApJSS] {10.1086/516591}, 172, 270

\bibitem[\protect\citeauthoryear{Cavanagh, Bekki  \& Groves}{Cavanagh
  et~al.}{2021}]{cavanagh2021}
Cavanagh M.~K.,  Bekki K.,   Groves B.~A.,  2021, \mn@doi [MNRAS]
  {10.1093/mnras/stab1552}, 506, 659

\bibitem[\protect\citeauthoryear{Cavanagh, Bekki, Groves  \& Pfeffer}{Cavanagh
  et~al.}{2022}]{cavanagh2022}
Cavanagh M.~K.,  Bekki K.,  Groves B.~A.,   Pfeffer J.,  2022, \mn@doi [MNRAS]
  {10.1093/mnras/stab3786}, 510, 5164

\bibitem[\protect\citeauthoryear{Cheng, {Huertas-Company}, Conselice,
  {Arag{\'o}n-Salamanca}, Robertson  \& Ramachandra}{Cheng
  et~al.}{2021}]{cheng2021}
Cheng T.-Y.,  {Huertas-Company} M.,  Conselice C.~J.,  {Arag{\'o}n-Salamanca}
  A.,  Robertson B.~E.,   Ramachandra N.,  2021, \mn@doi [MNRAS]
  {10.1093/mnras/stab734}, 503, 4446

\bibitem[\protect\citeauthoryear{Chollet et~al.}{Chollet
  et~al.}{2015}]{chollet2015}
Chollet F.,  et~al., 2015, Keras, \url{https://keras.io}

\bibitem[\protect\citeauthoryear{Coccato, {Fraser-McKelvie}, Jaff{\'e},
  Johnston, Cortesi  \& Pallero}{Coccato et~al.}{2022}]{coccato2022}
Coccato L.,  {Fraser-McKelvie} A.,  Jaff{\'e} Y.~L.,  Johnston E.~J.,  Cortesi
  A.,   Pallero D.,  2022, \mn@doi [MNRAS] {10.1093/mnras/stac1764}, 515, 201

\bibitem[\protect\citeauthoryear{Conselice}{Conselice}{2014}]{conselice2014}
Conselice C.~J.,  2014, \mn@doi [ARA\&A] {10.1146/annurev-astro-081913-040037},
  52, 291

\bibitem[\protect\citeauthoryear{D'Onofrio, Marziani  \& Buson}{D'Onofrio
  et~al.}{2015}]{donofrio2015}
D'Onofrio M.,  Marziani P.,   Buson L.,  2015, \mn@doi [Frontiers in Astronomy
  and Space Sciences] {10.3389/fspas.2015.00004}, 2

\bibitem[\protect\citeauthoryear{Deeley et~al.,}{Deeley
  et~al.}{2020}]{deeley2020}
Deeley S.,  et~al., 2020, \mn@doi [MNRAS] {10.1093/mnras/staa2417}, 498, 2372

\bibitem[\protect\citeauthoryear{Deeley, Drinkwater, Sweet, Bekki, Couch,
  Forbes  \& Dolfi}{Deeley et~al.}{2021}]{deeley2021}
Deeley S.,  Drinkwater M.~J.,  Sweet S.~M.,  Bekki K.,  Couch W.~J.,  Forbes
  D.~A.,   Dolfi A.,  2021, \mn@doi [MNRAS] {10.1093/mnras/stab2007}, 508, 895

\bibitem[\protect\citeauthoryear{Deger et~al.,}{Deger et~al.}{2018}]{deger2018}
Deger S.,  et~al., 2018, \mn@doi [ApJ] {10.3847/1538-4357/aaeb87}, 869, 6

\bibitem[\protect\citeauthoryear{Deng, Dong, Socher, Li, {Kai Li}  \& {Li
  Fei-Fei}}{Deng et~al.}{2009}]{deng2009}
Deng J.,  Dong W.,  Socher R.,  Li L.-J.,  {Kai Li}  {Li Fei-Fei} 2009, in 2009
  {{IEEE Conference}} on {{Computer Vision}} and {{Pattern Recognition}}.
  {IEEE}, {Miami, FL}, pp 248--255, \mn@doi{10.1109/CVPR.2009.5206848}

\bibitem[\protect\citeauthoryear{Desai et~al.,}{Desai et~al.}{2007}]{desai2007}
Desai V.,  et~al., 2007, \mn@doi [ApJ] {10.1086/513310}, 660, 1151

\bibitem[\protect\citeauthoryear{Diaz, Bekki, Forbes, Couch, Drinkwater  \&
  Deeley}{Diaz et~al.}{2018}]{diaz2018}
Diaz J.,  Bekki K.,  Forbes D.~A.,  Couch W.~J.,  Drinkwater M.~J.,   Deeley
  S.,  2018, \mn@doi [MNRAS] {10.1093/mnras/sty743}, 477, 2030

\bibitem[\protect\citeauthoryear{Dieleman, Willett  \& Dambre}{Dieleman
  et~al.}{2015}]{dieleman2015}
Dieleman S.,  Willett K.~W.,   Dambre J.,  2015, \mn@doi [MNRAS]
  {10.1093/mnras/stv632}, 450, 1441

\bibitem[\protect\citeauthoryear{Dom{\'i}nguez~S{\'a}nchez, {Huertas-Company},
  Bernardi, Tuccillo  \& Fischer}{Dom{\'i}nguez~S{\'a}nchez
  et~al.}{2018}]{dominguezsanchez2018}
Dom{\'i}nguez~S{\'a}nchez H.,  {Huertas-Company} M.,  Bernardi M.,  Tuccillo
  D.,   Fischer J.~L.,  2018, \mn@doi [MNRAS] {10.1093/mnras/sty338}, 476, 3661

\bibitem[\protect\citeauthoryear{Dom{\'i}nguez~S{\'a}nchez
  et~al.,}{Dom{\'i}nguez~S{\'a}nchez et~al.}{2019}]{dominguezsanchez2019}
Dom{\'i}nguez~S{\'a}nchez H.,  et~al., 2019, \mn@doi [MNRAS]
  {10.1093/mnras/sty3497}, 484, 93

\bibitem[\protect\citeauthoryear{Dom{\'i}nguez~S{\'a}nchez, Bernardi,
  Nikakhtar, {Margalef-Bentabol}  \& Sheth}{Dom{\'i}nguez~S{\'a}nchez
  et~al.}{2020}]{dominguezsanchez2020}
Dom{\'i}nguez~S{\'a}nchez H.,  Bernardi M.,  Nikakhtar F.,  {Margalef-Bentabol}
  B.,   Sheth R.~K.,  2020, \mn@doi [MNRAS] {10.1093/mnras/staa1364}, 495, 2894

\bibitem[\protect\citeauthoryear{Dressler}{Dressler}{1980}]{dressler1980}
Dressler A.,  1980, \mn@doi [ApJ] {10.1086/157753}, 236, 351

\bibitem[\protect\citeauthoryear{Dressler et~al.,}{Dressler
  et~al.}{1997}]{dressler1997}
Dressler A.,  et~al., 1997, \mn@doi [ApJ] {10.1086/304890}, 490, 577

\bibitem[\protect\citeauthoryear{Driver et~al.,}{Driver
  et~al.}{2009}]{driver2009}
Driver S.~P.,  et~al., 2009, \mn@doi [Astronomy and Geophysics]
  {10.1111/j.1468-4004.2009.50512.x}, 50, 5.12

\bibitem[\protect\citeauthoryear{{Eliche-Moral}, {Rodr{\'i}guez-P{\'e}rez},
  Borlaff, Querejeta  \& Tapia}{{Eliche-Moral} et~al.}{2018}]{eliche-moral2018}
{Eliche-Moral} M.~C.,  {Rodr{\'i}guez-P{\'e}rez} C.,  Borlaff A.,  Querejeta
  M.,   Tapia T.,  2018, \mn@doi [A\&A] {10.1051/0004-6361/201832911}, 617,
  A113

\bibitem[\protect\citeauthoryear{Fasano, Poggianti, Couch, Bettoni, Kjargaard
  \& Moles}{Fasano et~al.}{2000}]{fasano2000}
Fasano G.,  Poggianti B.~M.,  Couch W.~J.,  Bettoni D.,  Kjargaard P.,   Moles
  M.,  2000, \mn@doi [ApJ] {10.1086/317047}, 542, 673

\bibitem[\protect\citeauthoryear{Ferreira et~al.,}{Ferreira
  et~al.}{2022}]{ferreira2022}
Ferreira L.,  et~al., 2022, preprint
  (\href{https://arxiv.org/abs/2210.01110}{arXiv:2210.01110})

\bibitem[\protect\citeauthoryear{{Fraser-McKelvie}, {Arag{\'o}n-Salamanca},
  Merrifield, Tabor, Bernardi, Drory, Parikh  \&
  {Argudo-Fern{\'a}ndez}}{{Fraser-McKelvie} et~al.}{2018}]{fraser-mckelvie2018}
{Fraser-McKelvie} A.,  {Arag{\'o}n-Salamanca} A.,  Merrifield M.,  Tabor M.,
  Bernardi M.,  Drory N.,  Parikh T.,   {Argudo-Fern{\'a}ndez} M.,  2018,
  \mn@doi [MNRAS] {10.1093/mnras/sty2563}, 481, 5580

\bibitem[\protect\citeauthoryear{Ghosh, Urry, Wang, Schawinski, Turp  \&
  Powell}{Ghosh et~al.}{2020}]{ghosh2020}
Ghosh A.,  Urry C.~M.,  Wang Z.,  Schawinski K.,  Turp D.,   Powell M.~C.,
  2020, \mn@doi [ApJ] {10.3847/1538-4357/ab8a47}, 895, 112

\bibitem[\protect\citeauthoryear{Goodfellow, Bengio  \& Courville}{Goodfellow
  et~al.}{2016}]{goodfellow2016}
Goodfellow I.,  Bengio Y.,   Courville A.,  2016, Deep Learning.
Adaptive Computation and Machine Learning, {The MIT Press}, {Cambridge,
  Massachusetts}

\bibitem[\protect\citeauthoryear{Graham et~al.,}{Graham
  et~al.}{2018}]{graham2018}
Graham M.~T.,  et~al., 2018, \mn@doi [MNRAS] {10.1093/mnras/sty504}, 477, 4711

\bibitem[\protect\citeauthoryear{Gunn \& Gott}{Gunn \& Gott}{1972}]{gunn1972}
Gunn J.~E.,  Gott III J.~R.,  1972, \mn@doi [ApJ] {10.1086/151605}, 176, 1

\bibitem[\protect\citeauthoryear{Guo et~al.,}{Guo et~al.}{2011}]{guo2011}
Guo Q.,  et~al., 2011, \mn@doi [MNRAS] {10.1111/j.1365-2966.2010.18114.x}, 413,
  101

\bibitem[\protect\citeauthoryear{Harris et~al.,}{Harris
  et~al.}{2020}]{harris2020}
Harris C.~R.,  et~al., 2020, \mn@doi [Nature] {10.1038/s41586-020-2649-2}, 585,
  357

\bibitem[\protect\citeauthoryear{Hausen \& Robertson}{Hausen \&
  Robertson}{2020}]{hausen2020}
Hausen R.,  Robertson B.~E.,  2020, \mn@doi [ApJS] {10.3847/1538-4365/ab8868},
  248, 20

\bibitem[\protect\citeauthoryear{Haykin}{Haykin}{2009}]{haykin2009}
Haykin S.,  2009, Neural Networks and Learning Machines, 3rd edn.
{Pearson}, {New York}

\bibitem[\protect\citeauthoryear{Hinshaw et~al.,}{Hinshaw
  et~al.}{2013}]{hinshaw2013}
Hinshaw G.,  et~al., 2013, \mn@doi [ApJS] {10.1088/0067-0049/208/2/19}, 208, 19

\bibitem[\protect\citeauthoryear{Holden et~al.,}{Holden
  et~al.}{2009}]{holden2009}
Holden B.~P.,  et~al., 2009, \mn@doi [ApJ] {10.1088/0004-637X/693/1/617}, 693,
  617

\bibitem[\protect\citeauthoryear{Hubble}{Hubble}{1936}]{hubble1936}
Hubble E.~P.,  1936, Realm of the {{Nebulae}}.
{Yale University Press}

\bibitem[\protect\citeauthoryear{{Huertas-Company} et~al.,}{{Huertas-Company}
  et~al.}{2015}]{huertas-company2015}
{Huertas-Company} M.,  et~al., 2015, \mn@doi [ApJ]
  {10.1088/0004-637X/809/1/95}, 809, 95

\bibitem[\protect\citeauthoryear{Hunter}{Hunter}{2007}]{hunter2007}
Hunter J.~D.,  2007, \mn@doi [Computing in Science \& Engineering]
  {10.1109/MCSE.2007.55}, 9, 90

\bibitem[\protect\citeauthoryear{Ilbert et~al.,}{Ilbert
  et~al.}{2010}]{ilbert2010}
Ilbert O.,  et~al., 2010, \mn@doi [ApJ] {10.1088/0004-637X/709/2/644}, 709, 644

\bibitem[\protect\citeauthoryear{Ioffe \& Szegedy}{Ioffe \&
  Szegedy}{2015}]{ioffe2015}
Ioffe S.,  Szegedy C.,  2015, preprint
  (\href{https://arxiv.org/abs/1502.03167}{arXiv:1502.03167}

\bibitem[\protect\citeauthoryear{Jin, Dundar  \& Culurciello}{Jin
  et~al.}{2016}]{jin2016}
Jin J.,  Dundar A.,   Culurciello E.,  2016, preprint
  (\href{https://arxiv.org/abs/1511.06306}{arXiv:1511.06306})

\bibitem[\protect\citeauthoryear{Johnston, {Arag{\'o}n-Salamanca}  \&
  Merrifield}{Johnston et~al.}{2014}]{johnston2014}
Johnston E.~J.,  {Arag{\'o}n-Salamanca} A.,   Merrifield M.~R.,  2014, \mn@doi
  [MNRAS] {10.1093/mnras/stu582}, 441, 333

\bibitem[\protect\citeauthoryear{Johnston et~al.,}{Johnston
  et~al.}{2020}]{johnston2020}
Johnston E.~J.,  et~al., 2020, \mn@doi [MNRAS] {10.1093/mnras/staa2838}, 500,
  4193

\bibitem[\protect\citeauthoryear{Johnston et~al.,}{Johnston
  et~al.}{2022}]{johnston2022}
Johnston E.~J.,  et~al., 2022, \mn@doi [MNRAS] {10.1093/mnras/stac1447}, 514,
  6141

\bibitem[\protect\citeauthoryear{Just, Zaritsky, Sand, Desai  \& Rudnick}{Just
  et~al.}{2010}]{just2010}
Just D.~W.,  Zaritsky D.,  Sand D.~J.,  Desai V.,   Rudnick G.,  2010, \mn@doi
  [ApJ] {10.1088/0004-637X/711/1/192}, 711, 192

\bibitem[\protect\citeauthoryear{Kannappan, Guie  \& Baker}{Kannappan
  et~al.}{2009}]{kannappan2009}
Kannappan S.~J.,  Guie J.~M.,   Baker A.~J.,  2009, \mn@doi [AJ]
  {10.1088/0004-6256/138/2/579}, 138, 579

\bibitem[\protect\citeauthoryear{Kartaltepe et~al.,}{Kartaltepe
  et~al.}{2022}]{kartaltepe2022}
Kartaltepe J.~S.,  et~al., 2022, preprint
  (\href{https://arxiv.org/abs/2210.14713}{arXiv:2210.14713})

\bibitem[\protect\citeauthoryear{Kingma \& Ba}{Kingma \& Ba}{2014}]{kingma2014}
Kingma D.~P.,  Ba J.,  2014, preprint
  (\href{https://arxiv.org/abs/1412.6980}{arXiv:1412.6980})

\bibitem[\protect\citeauthoryear{Koekemoer et~al.,}{Koekemoer
  et~al.}{2007}]{koekemoer2007}
Koekemoer A.~M.,  et~al., 2007, \mn@doi [ApJS] {10.1086/520086}, 172, 196

\bibitem[\protect\citeauthoryear{Kormendy \& Kennicutt}{Kormendy \&
  Kennicutt}{2004}]{kormendy2004}
Kormendy J.,  Kennicutt R.~C.,  2004, \mn@doi [ARA\&A]
  {10.1146/annurev.astro.42.053102.134024}, 42, 603

\bibitem[\protect\citeauthoryear{Kova{\v c} et~al.,}{Kova{\v c}
  et~al.}{2010}]{kovac2010}
Kova{\v c} K.,  et~al., 2010, \mn@doi [ApJ]
  {10.1088/0004-637X/718/1/8610.48550/arXiv.0909.2032}, 718, 86

\bibitem[\protect\citeauthoryear{Laigle et~al.,}{Laigle
  et~al.}{2016}]{laigle2016}
Laigle C.,  et~al., 2016, \mn@doi [ApJS] {10.3847/0067-0049/224/2/24}, 224, 24

\bibitem[\protect\citeauthoryear{Larson, Tinsley  \& Caldwell}{Larson
  et~al.}{1980}]{larson1980}
Larson R.~B.,  Tinsley B.~M.,   Caldwell C.~N.,  1980, \mn@doi [ApJ]
  {10.1086/157917}, 237, 692

\bibitem[\protect\citeauthoryear{Laurikainen, Salo, Buta, Knapen, Speltincx  \&
  Block}{Laurikainen et~al.}{2006}]{laurikainen2006}
Laurikainen E.,  Salo H.,  Buta R.,  Knapen J.,  Speltincx T.,   Block D.,
  2006, \mn@doi [AJ] {10.1086/508810}, 132, 2634

\bibitem[\protect\citeauthoryear{Laurikainen, Salo, Buta, Knapen  \&
  Comer{\'o}n}{Laurikainen et~al.}{2010}]{laurikainen2010}
Laurikainen E.,  Salo H.,  Buta R.,  Knapen J.~H.,   Comer{\'o}n S.,  2010,
  \mn@doi [MNRAS] {10.1111/j.1365-2966.2010.16521.x}

\bibitem[\protect\citeauthoryear{LeCun, Bengio  \& Hinton}{LeCun
  et~al.}{2015}]{lecun2015}
LeCun Y.,  Bengio Y.,   Hinton G.,  2015, \mn@doi [Nature]
  {10.1038/nature14539}, 521, 436

\bibitem[\protect\citeauthoryear{Leauthaud et~al.,}{Leauthaud
  et~al.}{2007}]{leauthaud2007}
Leauthaud A.,  et~al., 2007, \mn@doi [ApJS] {10.1086/516598}, 172, 219

\bibitem[\protect\citeauthoryear{Marino, Bianchi, Rampazzo, Thilker, Annibali,
  Bressan  \& Buson}{Marino et~al.}{2011}]{marino2011}
Marino A.,  Bianchi L.,  Rampazzo R.,  Thilker D.~A.,  Annibali F.,  Bressan
  A.,   Buson L.~M.,  2011, \mn@doi [ApJ] {10.1088/0004-637X/736/2/154}, 736,
  154

\bibitem[\protect\citeauthoryear{Martin, Kaviraj, Hocking, Read  \&
  Geach}{Martin et~al.}{2020}]{martin2020}
Martin G.,  Kaviraj S.,  Hocking A.,  Read S.~C.,   Geach J.~E.,  2020, \mn@doi
  [MNRAS] {10.1093/mnras/stz3006}, 491, 1408

\bibitem[\protect\citeauthoryear{Massey, Stoughton, Leauthaud, Rhodes,
  Koekemoer, Ellis  \& Shaghoulian}{Massey et~al.}{2010}]{massey2010}
Massey R.,  Stoughton C.,  Leauthaud A.,  Rhodes J.,  Koekemoer A.,  Ellis R.,
   Shaghoulian E.,  2010, \mn@doi [MNRAS] {10.1111/j.1365-2966.2009.15638.x},
  401, 371

\bibitem[\protect\citeauthoryear{Masters et~al.,}{Masters
  et~al.}{2021}]{masters2021}
Masters K.~L.,  et~al., 2021, \mn@doi [MNRAS] {10.1093/mnras/stab2282}, 507,
  3923

\bibitem[\protect\citeauthoryear{{M{\'e}ndez-Abreu} et~al.,}{{M{\'e}ndez-Abreu}
  et~al.}{2018}]{mendez-abreu2018}
{M{\'e}ndez-Abreu} J.,  et~al., 2018, \mn@doi [MNRAS] {10.1093/mnras/stx2804},
  474, 1307

\bibitem[\protect\citeauthoryear{Mishra, Wadadekar  \& Barway}{Mishra
  et~al.}{2019}]{mishra2019}
Mishra P.~K.,  Wadadekar Y.,   Barway S.,  2019, \mn@doi [MNRAS]
  {10.1093/mnras/stz1621}, 487, 5572

\bibitem[\protect\citeauthoryear{Moran, Ellis, Treu, Smith, Rich  \&
  Smail}{Moran et~al.}{2007}]{moran2007}
Moran S.~M.,  Ellis R.~S.,  Treu T.,  Smith G.~P.,  Rich R.~M.,   Smail I.,
  2007, \mn@doi [ApJ] {10.1086/522303}, 671, 1503

\bibitem[\protect\citeauthoryear{Nair \& Abraham}{Nair \&
  Abraham}{2010}]{nair2010}
Nair P.~B.,  Abraham R.~G.,  2010, \mn@doi [ApJS]
  {10.1088/0067-0049/186/2/427}, 186, 427

\bibitem[\protect\citeauthoryear{Oesch et~al.,}{Oesch et~al.}{2010}]{oesch2010}
Oesch P.~A.,  et~al., 2010, \mn@doi [ApJ] {10.1088/2041-8205/714/1/L47}, 714,
  L47

\bibitem[\protect\citeauthoryear{Papovich, Dickinson, Giavalisco, Conselice  \&
  Ferguson}{Papovich et~al.}{2005}]{papovich2005}
Papovich C.,  Dickinson M.,  Giavalisco M.,  Conselice C.~J.,   Ferguson H.~C.,
   2005, \mn@doi [ApJ] {10.1086/429120}, 631, 101

\bibitem[\protect\citeauthoryear{Poggianti et~al.,}{Poggianti
  et~al.}{2001}]{poggianti2001}
Poggianti B.~M.,  et~al., 2001, \mn@doi [ApJ] {10.1086/323767}, 563, 118

\bibitem[\protect\citeauthoryear{Poggianti et~al.,}{Poggianti
  et~al.}{2009}]{poggianti2009}
Poggianti B.~M.,  et~al., 2009, \mn@doi [ApJ] {10.1088/0004-637X/697/2/L137},
  697, L137

\bibitem[\protect\citeauthoryear{Prieto et~al.,}{Prieto
  et~al.}{2013}]{prieto2013}
Prieto M.,  et~al., 2013, \mn@doi [MNRAS] {10.1093/mnras/sts065}, 428, 999

\bibitem[\protect\citeauthoryear{Querejeta, {Eliche-Moral}, Tapia, Borlaff,
  {Rodr{\'i}guez-P{\'e}rez}, Zamorano  \& Gallego}{Querejeta
  et~al.}{2015}]{querejeta2015}
Querejeta M.,  {Eliche-Moral} M.~C.,  Tapia T.,  Borlaff A.,
  {Rodr{\'i}guez-P{\'e}rez} C.,  Zamorano J.,   Gallego J.,  2015, \mn@doi
  [A\&A] {10.1051/0004-6361/201424303}, 573, A78

\bibitem[\protect\citeauthoryear{Rathore, Kumar, Mishra, Wadadekar  \&
  Bait}{Rathore et~al.}{2022}]{rathore2022}
Rathore H.,  Kumar K.,  Mishra P.~K.,  Wadadekar Y.,   Bait O.,  2022, \mn@doi
  [MNRAS] {10.1093/mnras/stac871}, 513, 389

\bibitem[\protect\citeauthoryear{Rizzo, Fraternali  \& Iorio}{Rizzo
  et~al.}{2018}]{rizzo2018}
Rizzo F.,  Fraternali F.,   Iorio G.,  2018, \mn@doi [MNRAS]
  {10.1093/mnras/sty347}, 476, 2137

\bibitem[\protect\citeauthoryear{Robertson et~al.,}{Robertson
  et~al.}{2022}]{robertson2022}
Robertson B.~E.,  et~al., 2022, preprint
  (\href{https://arxiv.org/abs/2208.11456}{arXiv:2208.11456})

\bibitem[\protect\citeauthoryear{Robotham et~al.,}{Robotham
  et~al.}{2014}]{robotham2014}
Robotham A. S.~G.,  et~al., 2014, \mn@doi [MNRAS] {10.1093/mnras/stu1604}, 444,
  3986

\bibitem[\protect\citeauthoryear{Saha \& Cortesi}{Saha \&
  Cortesi}{2018}]{saha2018}
Saha K.,  Cortesi A.,  2018, \mn@doi [ApJ] {10.3847/2041-8213/aad23a}, 862, L12

\bibitem[\protect\citeauthoryear{Schawinski et~al.,}{Schawinski
  et~al.}{2009}]{schawinski2009}
Schawinski K.,  et~al., 2009, \mn@doi [MNRAS]
  {10.1111/j.1365-2966.2009.14793.x}, 396, 818

\bibitem[\protect\citeauthoryear{Scoville et~al.,}{Scoville
  et~al.}{2007}]{scoville2007}
Scoville N.,  et~al., 2007, \mn@doi [ApJS] {10.1086/516585}, 172, 1

\bibitem[\protect\citeauthoryear{Simonyan \& Zisserman}{Simonyan \&
  Zisserman}{2014}]{simonyan2014}
Simonyan K.,  Zisserman A.,  2014, preprint
  (\href{https://arxiv.org/abs/1409.1556}{arXiv:1409.1556})

\bibitem[\protect\citeauthoryear{Somerville \& Dav{\'e}}{Somerville \&
  Dav{\'e}}{2015}]{somerville2015}
Somerville R.~S.,  Dav{\'e} R.,  2015, \mn@doi [ARA\&A]
  {10.1146/annurev-astro-082812-140951}, 53, 51

\bibitem[\protect\citeauthoryear{Szegedy, Vanhoucke, Ioffe, Shlens  \&
  Wojna}{Szegedy et~al.}{2015}]{szegedy2015}
Szegedy C.,  Vanhoucke V.,  Ioffe S.,  Shlens J.,   Wojna Z.,  2015, preprint
  (\href{https://arxiv.org/abs/1512.00567}{arXiv:1512.00567})

\bibitem[\protect\citeauthoryear{Tapia, {Carmen Eliche-Moral}, Aceves,
  {Rodr{\'i}guez-P{\'e}rez}, Borlaff  \& Querejeta}{Tapia
  et~al.}{2017}]{tapia2017}
Tapia T.,  {Carmen Eliche-Moral} M.,  Aceves H.,  {Rodr{\'i}guez-P{\'e}rez} C.,
   Borlaff A.,   Querejeta M.,  2017, \mn@doi [A\&A]
  {10.1051/0004-6361/201628821}, 604, A105

\bibitem[\protect\citeauthoryear{{The Astropy Collaboration} et~al.,}{{The
  Astropy Collaboration} et~al.}{2013}]{theastropycollaboration2013}
{The Astropy Collaboration} et~al., 2013, \mn@doi [A\&A]
  {10.1051/0004-6361/201322068}, 558, A33

\bibitem[\protect\citeauthoryear{Tous, Solanes  \& Perea}{Tous
  et~al.}{2020}]{tous2020}
Tous J.~L.,  Solanes J.~M.,   Perea J.~D.,  2020, \mn@doi [MNRAS]
  {10.1093/mnras/staa1408}, 495, 4135

\bibitem[\protect\citeauthoryear{Van~Kemenade et~al.,}{Van~Kemenade
  et~al.}{2022}]{vankemenade2022}
Van~Kemenade H.,  et~al., 2022, Python-Pillow/{{Pillow}}: 9.2.0, Zenodo,
  \mn@doi{10.5281/ZENODO.6788304}

\bibitem[\protect\citeauthoryear{{Vega-Ferrero} et~al.,}{{Vega-Ferrero}
  et~al.}{2021}]{vega-ferrero2021}
{Vega-Ferrero} J.,  et~al., 2021, \mn@doi [MNRAS] {10.1093/mnras/stab594}, 506,
  1927

\bibitem[\protect\citeauthoryear{Vogelsberger, Marinacci, Torrey  \&
  Puchwein}{Vogelsberger et~al.}{2020}]{vogelsberger2020}
Vogelsberger M.,  Marinacci F.,  Torrey P.,   Puchwein E.,  2020, \mn@doi
  [Nature Reviews Physics] {10.1038/s42254-019-0127-2}, 2, 42

\bibitem[\protect\citeauthoryear{Vulcani et~al.,}{Vulcani
  et~al.}{2011}]{vulcani2011}
Vulcani B.,  et~al., 2011, \mn@doi [MNRAS] {10.1111/j.1365-2966.2010.18182.x},
  413, 921

\bibitem[\protect\citeauthoryear{Walmsley et~al.,}{Walmsley
  et~al.}{2021}]{walmsley2021}
Walmsley M.,  et~al., 2021, \mn@doi [MNRAS] {10.1093/mnras/stab2093}, 509, 3966

\bibitem[\protect\citeauthoryear{Walmsley et~al.,}{Walmsley
  et~al.}{2022}]{walmsley2022}
Walmsley M.,  et~al., 2022, \mn@doi [MNRAS] {10.1093/mnras/stac525}, 513, 1581

\bibitem[\protect\citeauthoryear{Weaver et~al.,}{Weaver
  et~al.}{2022}]{weaver2022}
Weaver J.~R.,  et~al., 2022, \mn@doi [ApJS] {10.3847/1538-4365/ac3078}, 258, 11

\bibitem[\protect\citeauthoryear{Wei, Kannappan, Vogel  \& Baker}{Wei
  et~al.}{2010}]{wei2010}
Wei L.~H.,  Kannappan S.~J.,  Vogel S.~N.,   Baker A.~J.,  2010, \mn@doi [ApJ]
  {10.1088/0004-637X/708/1/841}, 708, 841

\bibitem[\protect\citeauthoryear{Weiss, Khoshgoftaar  \& Wang}{Weiss
  et~al.}{2016}]{weiss2016}
Weiss K.,  Khoshgoftaar T.~M.,   Wang D.,  2016, \mn@doi [Journal of Big Data]
  {10.1186/s40537-016-0043-6}, 3, 9

\bibitem[\protect\citeauthoryear{Willett et~al.,}{Willett
  et~al.}{2013}]{willett2013}
Willett K.~W.,  et~al., 2013, \mn@doi [MNRAS] {10.1093/mnras/stt1458}, 435,
  2835

\bibitem[\protect\citeauthoryear{Williams, Bureau  \& Cappellari}{Williams
  et~al.}{2010}]{williams2010}
Williams M.~J.,  Bureau M.,   Cappellari M.,  2010, \mn@doi [MNRAS]
  {10.1111/j.1365-2966.2010.17406.x}, 409, 1330

\bibitem[\protect\citeauthoryear{Wilman \& Erwin}{Wilman \&
  Erwin}{2012}]{wilman2012}
Wilman D.~J.,  Erwin P.,  2012, \mn@doi [ApJ] {10.1088/0004-637X/746/2/160},
  746, 160

\bibitem[\protect\citeauthoryear{Wilman, Oemler, Mulchaey, McGee, Balogh  \&
  Bower}{Wilman et~al.}{2009}]{wilman2009}
Wilman D.~J.,  Oemler Jr. A.,  Mulchaey J.~S.,  McGee S.~L.,  Balogh M.~L.,
  Bower R.~G.,  2009, \mn@doi [ApJ] {10.1088/0004-637X/692/1/298}, 692, 298

\bibitem[\protect\citeauthoryear{{van den Bergh}}{{van den
  Bergh}}{2009}]{vandenbergh2009}
{van den Bergh} S.,  2009, \mn@doi [ApJ] {10.1088/0004-637X/702/2/1502}, 702,
  1502

\bibitem[\protect\citeauthoryear{{van der Wel} et~al.,}{{van der Wel}
  et~al.}{2007}]{vanderwel2007}
{van der Wel} A.,  et~al., 2007, \mn@doi [ApJ] {10.1086/521783}, 670, 206

\makeatother
\end{thebibliography}

\appendix

\section{Noise Augmentation and Transfer Learning}

\begin{figure}
    \centering
    \includegraphics[scale=0.59]{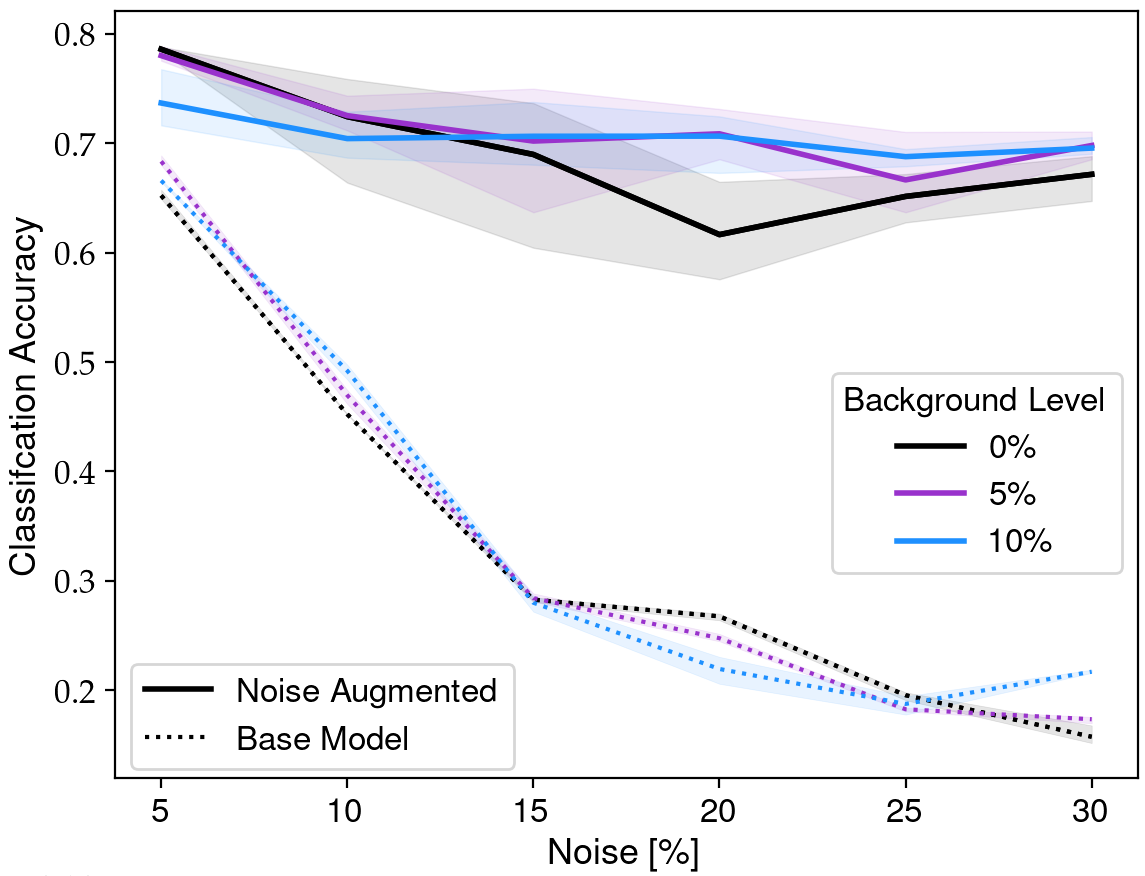}
    \caption{Plot of the classification accuracy for models classifying a test dataset at different, fixed levels of artificial noise and background. Shaded regions denote the full range of accuracies and losses for each of the three trials.}
    \label{fig:noiseacc}
\end{figure}
    
Transfer learning is a general technique for domain adaptation, in which an existing model trained on one domain (or dataset) can instead be adapted to create a new model applicable to a different, yet related, domain (see \citealt{weiss2016} for a review). The key motivation behind this approach is to leverage the pre-existing capabilities of existing models. Common present-day examples include taking large pretrained models such as Inception \citep{szegedy2015} that have been trained on very large, general-purpose datasets (such as ImageNet \citealt{deng2009}) and then adapting these for image classification tasks, such galaxy classification \citep{ackermann2018}. Transfer learning has also been utilised in astronomy to adapt models to classify images in different astronomical surveys using solely the known morphologies from the original, base survey \citep{dominguezsanchez2019}. Not only is this technique efficient, but it is especially useful when there is insufficient labelled data to effectively train a full model from scratch. This work utilises transfer learning together with data augmentation to classify high-redshift COSMOS images.

\subsection{Data Augmentation with Artificial Noise}

\begin{figure*}
    \centering
    \includegraphics[scale=0.495]{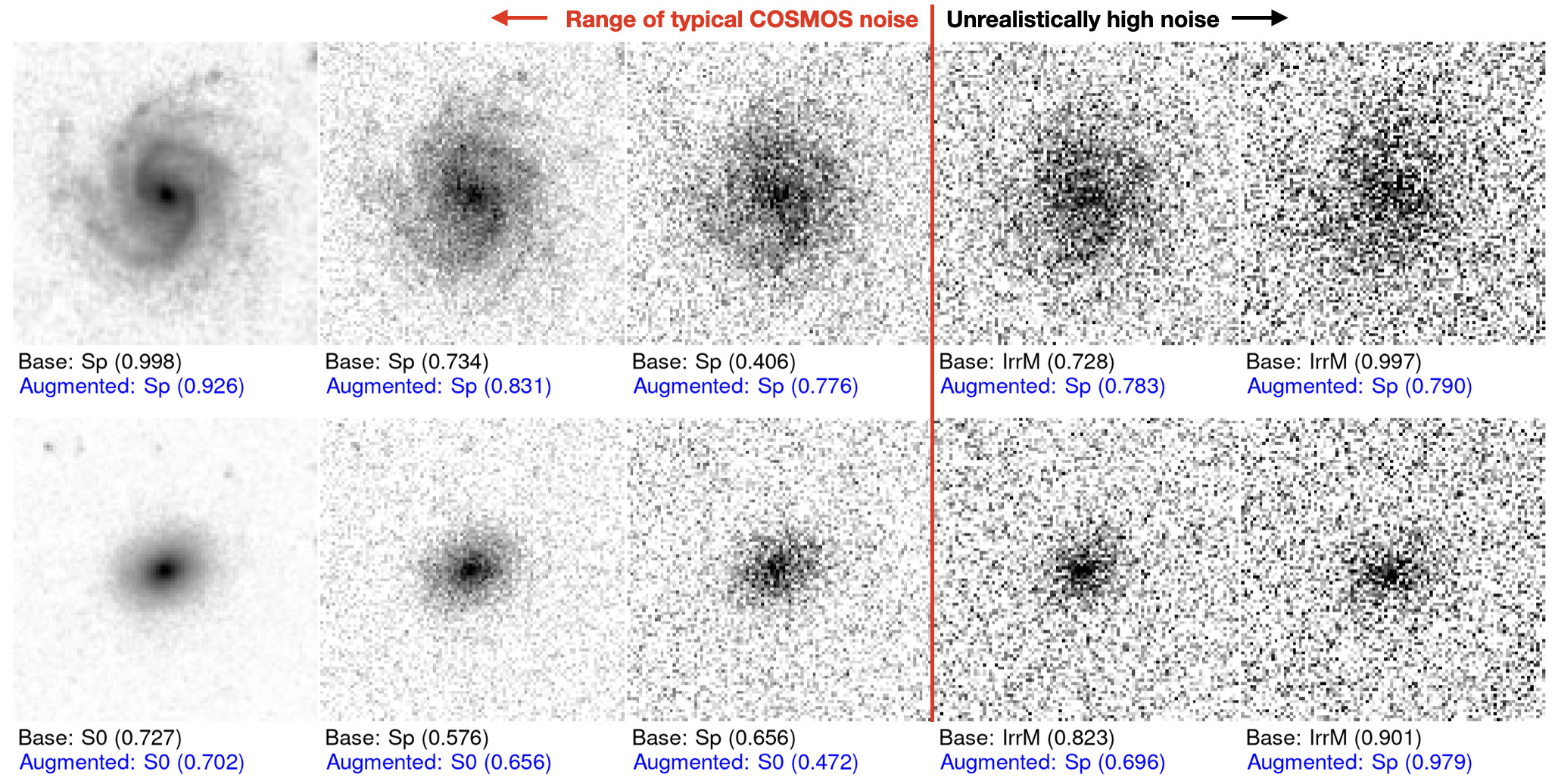}
    \caption{{Morphological classifications for an example spiral and S0 galaxy from the NA10 test set when subject to increasing levels of artificial Gaussian noise. Classifications are obtained using the original 4-class base model, and the final noise-augmented 4-class ensemble model. For first column shows the original image with no noise, while the next columns show the same image with noise levels of 15\%, 30\%, 40\% and 50\%, with the vertical red line marking the limit of typical COSMOS noise levels. Classification confidences are shown in parentheses.}}
    \label{fig:noiselevelexample}
\end{figure*}
    
There are several challenges that arise when shifting to a different observation space. Foremost among them is the issue of noise, resulting in degraded image quality and ultimately impacting the ability to classify the morphologies of more distant galaxies. Noise is one of several issues, including PSF smoothing and k-correction, that impact the ability to classify high-redshift samples, especially for distant galaxies. This work is focused solely on addressing noise, in particular by adapting our pretrained models to classify noisy images. { We achieve this through adding artificial noise to our existing training data of SDSS images in order to replicate the characteristic noise of the COSMOS images \citep{koekemoer2007,leauthaud2007}. This allows us to simply adapt the model using the existing images and their labels (see also \citealt{vega-ferrero2021}).} This noise is applied on-demand during training through the use of a custom \textsc{Keras} preprocessing layer, which applies a random level of additive Gaussian noise to the input images per minibatch for every epoch of training.
    
In general, corrupting real-valued inputs with additive Gaussian noise is a known regularisation technique that is especially effective at {improving model robustness and} reducing a model's susceptibility to adversarial inputs, i.e. inputs with random perturbations intended to ``fool'' the network \citep{jin2016}. For our purposes, we wish for our models to remain effective at classifying noisy images. Our noise augmentation process is tightly controlled so that the artificial noise is representative of the typical noise of the images in our COSMOS dataset.
    
The exact procedure involves adding Gaussian noise directly to the normalised pixel values. Importantly, the amount of noise added is not fixed; instead, the mean and standard deviation for the Gaussian noise is chosen uniformly within a range of predetermined bounds. To determine suitable bounds, we first analysed the background noise of random batches of COSMOS images at varying redshifts, obtaining approximate Gaussian fits. These bounds were then further refined through a visual comparison of the artificially noisy SDSS images with the actual COSMOS images. The values we obtained range from between -0.02 to 0.1 inclusive for the mean (background) and between 0 and 0.3 inclusive for the standard deviation (noise). This compares well with background levels of 0 to 10\% of the peak flux, and a noise threshold of 0.1\% to 40\% of the peak flux from \citet{leauthaud2007}. Since these values are chosen uniformly there is no bias for morphology. With this custom noise augmentation layer, it is possible to apply arbitrary levels of noise within these bounds to a given SDSS image. This is important firstly to avoid overfitting (by ensuring that images are not all given identical levels of noise), and secondly to ensure that the augmented images collectively reflect the range of inherent image qualities and signal-to-noise values across our COSMOS dataset. Figure \ref{fig:noiseaug-example} shows an example of this procedure in action. After the noise is added, the resulting pixel values are clipped to between 0 and 1 inclusive. We stress that such noise is deliberately artificial and is not intended to be a 100\% accurate reflection of the physical noise levels. The primary purpose of the noise augmentation is to help adapt the CNN to better classify degraded images.

\subsection{The Impact of Artificial Noise}
    
To analyse the impact of this noise augmentation on the overall classification accuracies of our CNNs, we tested the procedure using different, {discrete} Gaussian noise and background levels. We performed three trials for each fixed noise and background level. In each trial, an augmented model is both trained and tested with noise-augmented images. To compare performance, the base model is also tested with the same noise-augmented images as with the augmented model. Figure \ref{fig:noiseacc} shows a dramatic difference between the initial models and the noise-augmented models when tasked with classifying the augmented images. The initial models struggle to classify noisy images, with accuracies no better than random even at relatively modest levels of noise. The noise-augmented models have superior performance, {with accuracies as high as 70\% at the maximum typical noise level of 30\%.} The noise augmentation process is also highly variable, with the accuracies of the three individual trials differing by around 10\%. Background levels have little impact on accuracy.

{We further tested the models using unrealistically high levels of noise; that is, noise levels higher than the bounds we obtained from analysing the COSMOS images, and henceforth higher than the levels used to train the noise-augmented models. Figure~\ref{fig:noiselevelexample} compares the classifications of the original 4-class base models with those of the final ensemble of augmented models (see section below). In the case of the example spiral galaxy, both models correctly identify its morphological type throughout the range of typical COSMOS noise levels, however the classification confidence for the augmented ensemble is higher. In the case of the example S0 galaxy, the base model incorrectly classifies the noisy images as spirals, while the augmented model correctly identifies them as an S0 (albeit with relatively lower confidence levels). Both models degrade with unrealistically high levels of noise.} It is nevertheless clear that models not explicitly adapted to classify noisy images will struggle.

\subsection{Transfer Learning}

To commence transfer learning, we first require pretrained base models. These are the 3-class and 4-class models trained on SDSS images from the NA10 catalogue with the new architecture as described in the previous subsection. There are several methods for conducting transfer learning, such as partially or completely freezing layers of a model (essentially keeping the existing parameters as they are) while tweaking other layers, or fine-tuning the entire model. We utilise the latter, fine-tuning approach. We instantiate the model with its existing weights, then (re)train it with the new noise augmentation layers, albeit with a much lower learning rate (reduced by a factor of 10 to $8 \times 10^{-5}$). {Note that the model architecture and total number of trainable parameters remain unchanged.} The training data also remains identical, including the dedicated test set used for evaluation, which is also subject to the noise augmentation process to ensure fair evaluation. {Since the noise augmentation process is highly variable,} we train ensembles of five 3-class models and five 4-class models. The final classification of a given image is obtained through averaging the output probabilities across each of the five individual models. The predicted class is henceforth defined as the class with the highest mean output probability. We trained these models using the SDSS images from the NA10 dataset, with 80\% of the sample constituting the training set, and 20\% set aside as a test set for final evaluation. To fairly evaluate the noise augmented models, the images in the test set are also augmented with a random amount of noise.

\bsp	
\label{lastpage}
\end{document}